\documentclass[%
 reprint,
nofootinbib,
 amsmath,amssymb,
 prl,
floatfix,
]{revtex4-2}
\usepackage{amsmath}
\usepackage{amssymb}
\usepackage[tight]{subfigure}
\usepackage{enumitem}
\usepackage{soul}
\usepackage{cancel}
\usepackage{multirow}
\usepackage{tikz}
\usepackage{pifont}
\usepackage{xspace}
\usepackage{comment}
\usepackage{physics}
\usepackage{bbold}
\usepackage{graphicx}
\usepackage{dcolumn}
\usepackage{bm}
\usepackage{hyperref}
\usepackage[capitalise]{cleveref}

\begin{document}

\preprint{APS/123-QED}

\title{Robust Oscillations and Edge Modes in Nonunitary Floquet Systems}
\author{Vikram Ravindranath}
\author{Xiao Chen}%
\affiliation{Department of Physics, Boston College, Chestnut Hill, MA 02467, USA}

\date{\today}

\begin{abstract}
We explore oscillatory behaviour in a family of periodically driven spin chains which are subject to a weak measurement followed by post-selection. We discover a transition to an oscillatory phase as the strength of the measurement is increased. By mapping these spin chains to free fermion models, we find that this transition is reflected in the opening of a gap in the imaginary direction. Interestingly, we find a robust, purely real, edge $\pi$-mode in the oscillatory phase. We establish a correspondence between the complex bulk spectrum and these edge modes. These oscillations are numerically found to be stable against interactions and disorder.
\end{abstract}

\maketitle

\textit{Introduction} -- Recent years have witnessed a growing interest in the monitored many-body quantum dynamics. It has been shown that there exists a generic entanglement phase transition in a unitary quantum dynamics subject to continuous monitoring \cite{skinner2019measurement,Li_2018,Chan_2019,choi2020quantum,gullans2020dynamical}. By varying the monitoring strength, the individual quantum trajectory changes from a highly entangled volume-law phase to a disentangled area-law phase. Besides this  phase transition, monitoring quantum dynamics can generate novel quantum phases, which exhibit quantum criticality or even host quantum orders \cite{Chen_2020,Alberton_2021,Sang_2021,Lavasani_2021,Ippoliti_2021,Basu_2022,Jian_2021}. Here the order can be conventional order or topological order, and is determined by the form of the measurement operator.

Most of these studies are focused on the static order in the steady state. In this paper, we explore quantum ordered phases with oscillatory behavior in a monitored qubit system. We investigate this behavior in a periodically driven non-unitary circuit. We show that the steady state can exhibit persistent oscillations between two ordered phases. Moreover, these oscillations break the discrete time-translation symmetry of the underlying dynamics, similar to time crystals which have been observed in disordered Floquet many-body localized systems \cite{PhysRevLett.116.250401,PhysRevLett.117.090402,mi2022time,doi:10.1126/science.abk0603}. In our model, the quantum order is protected by local \lq\lq{}forced" measurements that prefer specific ordered configurations. Applied local unitaries flip between these ordered configurations, leading to oscillations.

In systems which can be mapped to models of free fermions, we demonstrate that such an oscillation behavior is due to a non-Hermitian analog of Majorana zero modes. Such an idea has been used to understand the ground state degeneracy in the Ising spin chain\cite{Kitaev_2001}. It has further been employed to understand  \lq\lq{}(almost) strong modes" that exhibit long coherence times in various static and driven Hermitian systems\cite{Kemp_2017,PhysRevB.99.205419}. In our model, the zero mode exists in an {\it imaginary} gap in the spectrum, is localized on the boundary and anti-commutes with the evolution operator, resulting in persistent oscillation behavior. 

\textit{Non-unitary Floquet Dynamics} --
The dynamics of periodically-driven systems over one period $T$ is governed by the \textit{Floquet} operator $\hat{V}$ \cite{reichl2021transition}. Analogous to unitary dynamics, in the non-unitary case, the dynamics of the quantum state is given by the repeated application of $\hat{V}$ followed by an explicit normalization of the state,
\begin{align}
    \label{eq:erule}
    |\psi(NT)\rangle= \frac{\hat{V}(NT)\ket{\psi_0}}{\norm{\hat{V}(NT)\ket{\psi_0}}}=\frac{(\hat{V})^N\ket{\psi_0}}{\norm{(\hat{V})^N\ket{\psi_0}}},
\end{align}
where $T$ has been set to 1. To understand the properties of the steady states -- in the limit $N\to\infty$ -- we need to analyze the spectrum of the $\hat{V}$ operator.

It is also convenient to define an effective non-Hermitian Hamiltonian $\hat{H}_F$ by expressing $\hat{V}$ as $e^{-i\hat{H}_F}$.   
We denote the (complex) eigenvalues of $\hat{H}_F$ by $\qty{E_n}$, their corresponding right eigenstates by $\qty{\ket{E_n}}$ and order them such that $\Im{E_j}\geq\Im{E_{j+1}}$. A generic initial state can be expressed as
\begin{align}
\ket{\psi_0} = \sum c_j\ket{E_j}.
\end{align}
Under time evolution, the unnormalized state
\begin{align}
\ket{\psi(NT)} = \sum_j c_j e^{-iE^R_jN} e^{E^I_jN} \ket{E_j},
\end{align}
where $E^{R(I)}_j$ denotes the real (imaginary) part of $E_j$. Clearly, as $N\to\infty$, the quantum state approaches $|E_1\rangle$ with the largest $E_1^I$, provided that $c_1\neq0$.

In principle, a degeneracy in the imaginary direction can emerge so that $E^I_1 = E^I_2 = \dots = E^I_{N_S}$ for some $N_S\geq2$. Generic initial states then do not evolve to a single final state; instead, they continue to evolve, even at late times, in the subspace spanned by $\qty{\ket{E_1},\ket{E_2},\dots,\ket{E_{N_S}}}$. If the real parts $E^R_j$ are uniformly separated so that $E^R_j \equiv E_0 + j\frac{2\pi}{N_S}$ the steady states can exhibit periodic behavior, with period $N_S$. In this paper, we focus on $N_S=2$.

\textit{Free Fermions \& \lq\lq{}Zero\rq\rq{} Modes} -- When the Hamiltonian $\hat{H}$ can be mapped to a system of non-interacting fermions, the many-particle spectrum of $\hat{H}$ can be built up by independently filling in the different single-particle energy levels. Since the many-particle energies are the sums of the energies  of the occupied levels, a 2 dimensional steady state subspace manifests in the presence of a single-particle mode $c^\dagger_0$, the imaginary part of whose eigenvalue $\epsilon_0$ is 0. We term such modes that exist in the imaginary gap of the spectrum $i0$ \textit{modes}, to distinguish them from the familiar zero modes that exist in a real gap. Many-body eigenstates can be grouped into pairs that differ solely in the occupation of $c^\dagger_0$. States in a pair have energies with the same imaginary part, and real parts offset by $\epsilon_0$. 

With this picture in mind, given a $\hat{V}$ which describes non-interacting fermions, the many-body spectrum of $\hat{V}$ is now obtained from the product of its single-particle eigenvalues. A single-particle mode $c^\dagger_0$ with the property that $\hat{V}c^\dagger_0 = -c^\dagger_0 \hat{V} \qty(= e^{-i\pi} c^\dagger_0 \hat{V})$ generates a similar pairing of many-body states which differ in the occupation of $c^\dagger_0$, thereby having eigenvalues with the same absolute value, but differing in sign.

\textit{Model and setup} -- We consider a system of $L$ spins subject to periodic, non-unitary driving, described by a Floquet operator $\hat{V}$. We study operators $\hat{V}$ which can be written as a composition of an imaginary time evolution $\hat{U}_I$ and a unitary operator $\hat{U}_R$. We also define the non-hermitian Floqet Hamiltonian $\hat{H}_F$. We consider a specific form for $\hat{U}_I$ and $\hat{U}_R$, as shown below:
\begin{equation}
    \begin{aligned}
        \hat{V} =& \hat{U}_I \hat{U}_R\\
        \hat{U}_I =& e^{\beta \sum\limits_j \hat{Z}_j \hat{Z}_{j+1}}\\
        \hat{U}_R =& e^{-i \sum\limits_j J_{zz,j}\hat{Z}_j \hat{Z}_{j+1}}e^{-i \sum\limits_j J_{xx,j} \hat{X}_j \hat{X}_{j+1}} e^{-i h_y\sum\limits_j Y_{j}}\\
        \hat{H}_F &\equiv i \log{\hat{V}}
    \end{aligned}
    \label{eq:vdefine}
\end{equation}

$\hat{X}_j, \hat{Y}_j$ and $\hat{Z}_j$ refer to the Pauli operators acting non-trivially only on the spin at site $j$. The various parameters $\beta, J_{zz,j}, J_{xx,j}$ and $h_j$ are all real. $\hat{U}_I$ can be interpreted as a forced measurement with $\beta$ being the strength of the measurement. $\hat{U}_R$ is composed of unitaries that describe nearest-neighbor $XX$ and $ZZ$ couplings, and a pulse which rotates each spin by $2h_y$ about the $Y$-axis. The operator $\hat{V}$ has a $\mathbb{Z}_2$ symmetry represented by the Parity operator $P=\prod\limits_j \hat{Y}_j$, which represents a simultaneous $\pi$ rotation of every spin about the $Y-$axis. The time evolution proceeds according to \cref{eq:erule}.

In the simplest case, where the pulses are near-perfect $\pi-$rotations about the $Y-$ axis, the nearest-neighbor couplings $J_{xx} = J_{zz} = 0$, and the measurement strength $\beta\to\infty$, we expect to see oscillations between the two ordered phases $\ket{\uparrow\uparrow\uparrow\cdots}$ and $\ket{\downarrow\downarrow\downarrow\cdots}$. Our objective is to study the consequences of moving away from this fine tuned limit and examine if there exists a phase with finite $\beta$ in which oscillations are present.


\begin{figure}
    \raggedright
    \includegraphics[width=0.475\textwidth]{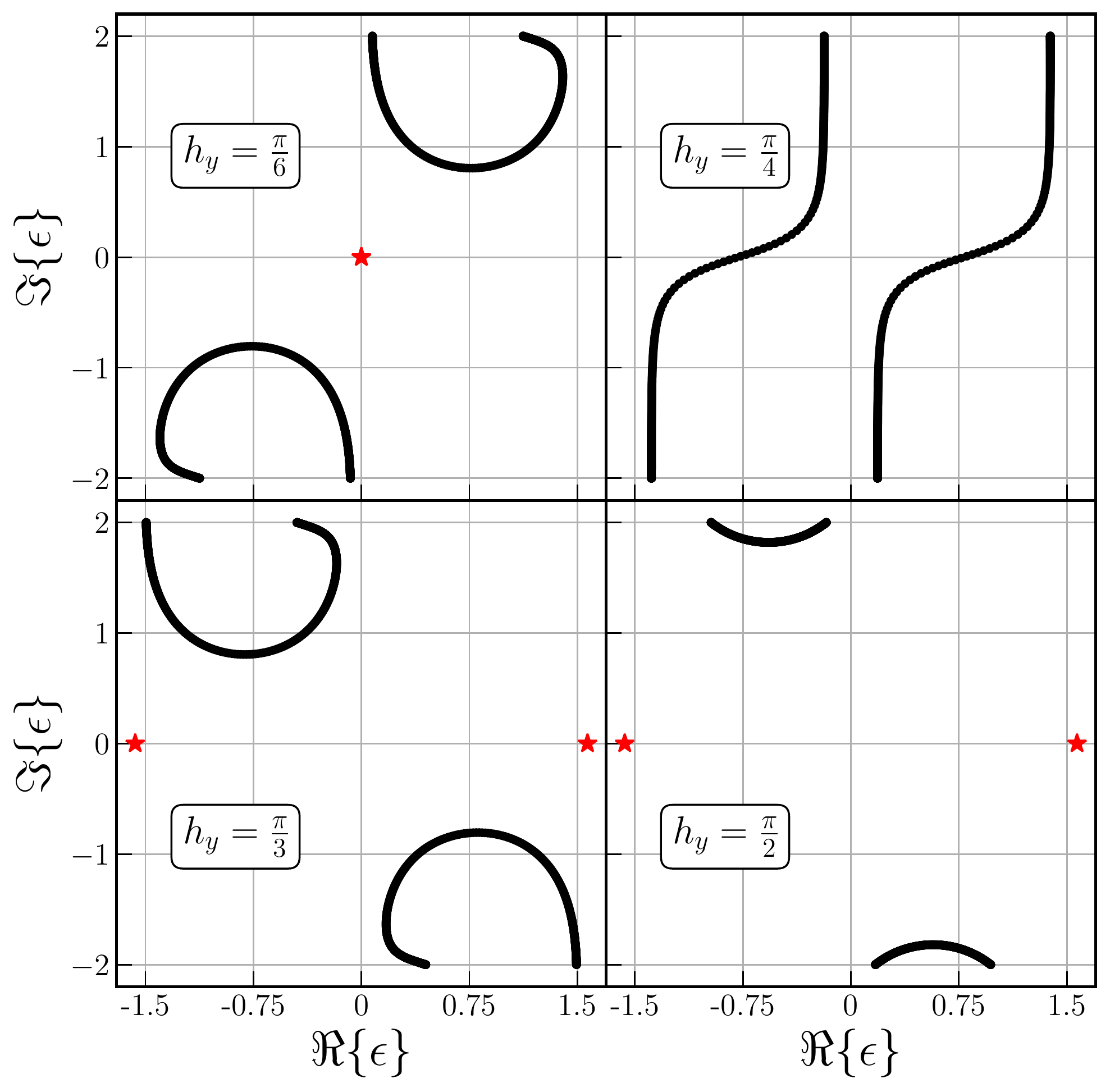}
    \caption{Complex spectrum of $\hat{H}_F$ given in \cref{eq:vdefine} with $L=1000$, $J_{xx} = 0.4, J_{zz}=1$ and $\beta=2$ with open boundary conditions for various $h_y$. The imaginary gap closes and reopens with the presence of a $\pi-$ splitting between the $i0$ modes.}
    \label{fig:spect1}
\end{figure}

\textit{Bulk Spectrum} -- We begin by studying the single particle spectrum of the operator $\hat{V}$ with a fixed $\beta$. The corresponding non-interacting Hamiltonian $H_F$ has the form $\hat{H}_F=\frac{1}{4}\sum_{i,j}\gamma_i \mathcal{H}_{ij} \gamma_j$, with $\mathcal{H}$ a complex, antisymmetric $2L\cross2L$ matrix and $\gamma_i$ being majorana fermion operators. This Hamiltonian can be diagonalized analogously to a Hermitian free fermion system \cite{Guo_2021}, albeit now with a complex spectrum and complex majorana-like operators ${g_j}$.

\begin{equation}
    \hat{H}_F = \frac{i}{2}\sum_{j=1}^L \epsilon_j g_{2j-1}g_{2j}.
\end{equation}

Since the $\hat{V}$ (hence also, $\hat{H}_F$) that we consider does not conserve particle number, we plot the quasi-energy spectrum in pairs of $\pm\frac{\epsilon}{2}$, where the $+(-)$ corresponds to a single particle state being occupied (unoccupied), resulting in $2L$ points being presented on each plot. A many-body eigenstate of $\hat{V}$ is determined by choosing one mode in each of the $L$ pairs. The process of obtaining and diagonalizing $\hat{H}_F$ is detailed in \cite{suppMat}.

We first focus on the regime where $\beta$ is large. Consequently, the spectrum is gapped for a wide range of $h_y$. As $h_y$ is varied, an eigenvalue gap closes and reopens in the imaginary direction. With open boundary conditions, $i0$ modes are present on either side of the gap closing. When $h_y$ is small, the modes are degenerate. However, when $h_y$ is tuned through the reopening of the imaginary gap, a $\pi$ splitting between the real parts of the $i0$ modes emerges, as shown in \cref{fig:spect1}. This is accompanied by the presence of oscillations in the steady state. The transition that we observe concerns the development of this robust splitting of $\pi$ in the real values -- not merely the presence -- of the $i0$ modes.

For the case where $J_{xx} = J_{zz} = 0$, the bulk spectrum of $H_F$, $\epsilon(k)$ can be obtained analytically.
\begin{equation}
    \begin{aligned}
        \epsilon(k) &= \frac{i}{2}\log(z(k) \pm i\sqrt{1-(z(k))^2})\\
        z(k) &= \cosh{2\beta}\cos(2h_y) + i\sinh(2\beta)\sin(2h_y)\cos(k)
    \end{aligned}
\end{equation}

The gap closing in the imaginary direction corresponds to $|z \pm i\sqrt{1-z^2}|=1$. This can only happen at $k=\frac{\pi}{2}$, requiring the condition
\begin{equation}
\label{eq:crit}
 |\cosh(2\beta)\cos(2h_y)|>1
\end{equation}
for the gap to remain open. The phases with and without a $\pi$-splitting between the real parts of the $i0$ modes' energies correspond to $\cosh(2\beta)\cos(2h_y) < -1$ and $\cosh(2\beta)\cos(2h_y) > 1$, respectively.

\textit{Edge \texorpdfstring{$i0$}{i0} Modes} -- We now analyze the $i0$ modes and their real-space distribution in detail. We begin by delineating the role of an $i0$ mode. When the imaginary time evolution part of $\hat{V}$ is sufficiently strong, the steady state is superposition of two states $\qty{|1\rangle,|2\rangle}$ which have opposite parity and are the right eigenvectors of $\hat{V}$. These are degenerate, in the sense that $\hat{V}\ket{i} = \lambda_i\ket{i}$ for $i=1,2$, $|\lambda_1|=|\lambda_2|$ and $|\lambda_1|>|\lambda_j|$ for all other eigenstates $\ket{j} \left(j\neq1,2\right)$ of $\hat{V}$. An $i0$ mode is an operator $\hat{F}$ that can toggle between $\ket{1}$ and $\ket{2}$.

Since it toggles between states of different parity, $\hat{F}$ anticommutes with $\hat{P}$. Further, since $|\lambda_1| = |\lambda_2|$, $\hat{V}\hat{F} = e^{i\theta}\hat{F}\hat{V}$, with $\theta$ real. This can be seen from the effect of $\hat{V}$ on a superposition of $|1\rangle, |2\rangle$, since if $\theta$ were complex, there would only be one steady state.
\begin{align}
    \hat{V}|\psi\rangle&=\hat{V} (a|1\rangle+b|2\rangle)=\hat{V} (a|1\rangle+b\hat{F}|1\rangle) \nonumber\\
    &=(a\hat{V}|1\rangle + e^{i\theta} b\hat{F}\hat{V}|1\rangle)= \lambda_1(a|1\rangle + e^{i\theta} b |2\rangle).
\end{align}

Further, since this work considers models with a 2 dimensional steady-state space, we must have $\hat{F}^2\ket{1} = e^{2i\theta}\ket{1}$, and $\hat{F}^2\ket{1} = \hat{F}\ket{2} = \ket{1}$, implying $\theta = 0$ or $\pi$. In the case where $\theta =\pi$, oscillations with twice the period of the driving are observed, and $\hat{V}$ and $\hat{F}$ anticommute. There are two $i0$ modes $\hat{F}_{(L)}$ and $\hat{F}_{(R)}$, localized on the left and right boundaries, respectively. The localization of these $i0$ modes guarantees the double degeneracy of the steady states in the thermodynamic limit. For instance, in the limit where $\beta\to\infty$ and $h_y\approx\frac{\pi}{2}$, the steady states are $\ket{n} = \frac{1}{\sqrt{2}}\qty(\ket{\uparrow\uparrow\uparrow\cdots} + (-1)^n i^L\ket{\downarrow\downarrow\downarrow\cdots})$ for $n=1,2$. The role of $\hat{F}_{L(R)}$ is played by $\hat{Z}_{1(L)}$, and both $\hat{Z}_{1,L}$ anticommute with $\hat{V}$.

Summarizing, the $i0$ mode satisfies \begin{enumerate}
    \item $\acomm{\hat{F}}{\hat{P}} = 0$
    \item $\acomm{\hat{F}}{\hat{V}} \to 0 $ as $L \to \infty$
    \item $\hat{F}$ decays exponentially into the bulk
    \item $\hat{F} = \sum\limits_{j=1}^L v_{2j-1} a_j + v_{2j} b_j$ in free fermion systems
\end{enumerate}

Additionally, we require that $\hat{F}^2 \propto \mathbb{1}$, since the steady state space is 2 dimensional. This condition is trivially satisfied by the ansatz (4) for free fermion systems.

\begin{figure}
    \centering
    \includegraphics[width=0.45\textwidth]{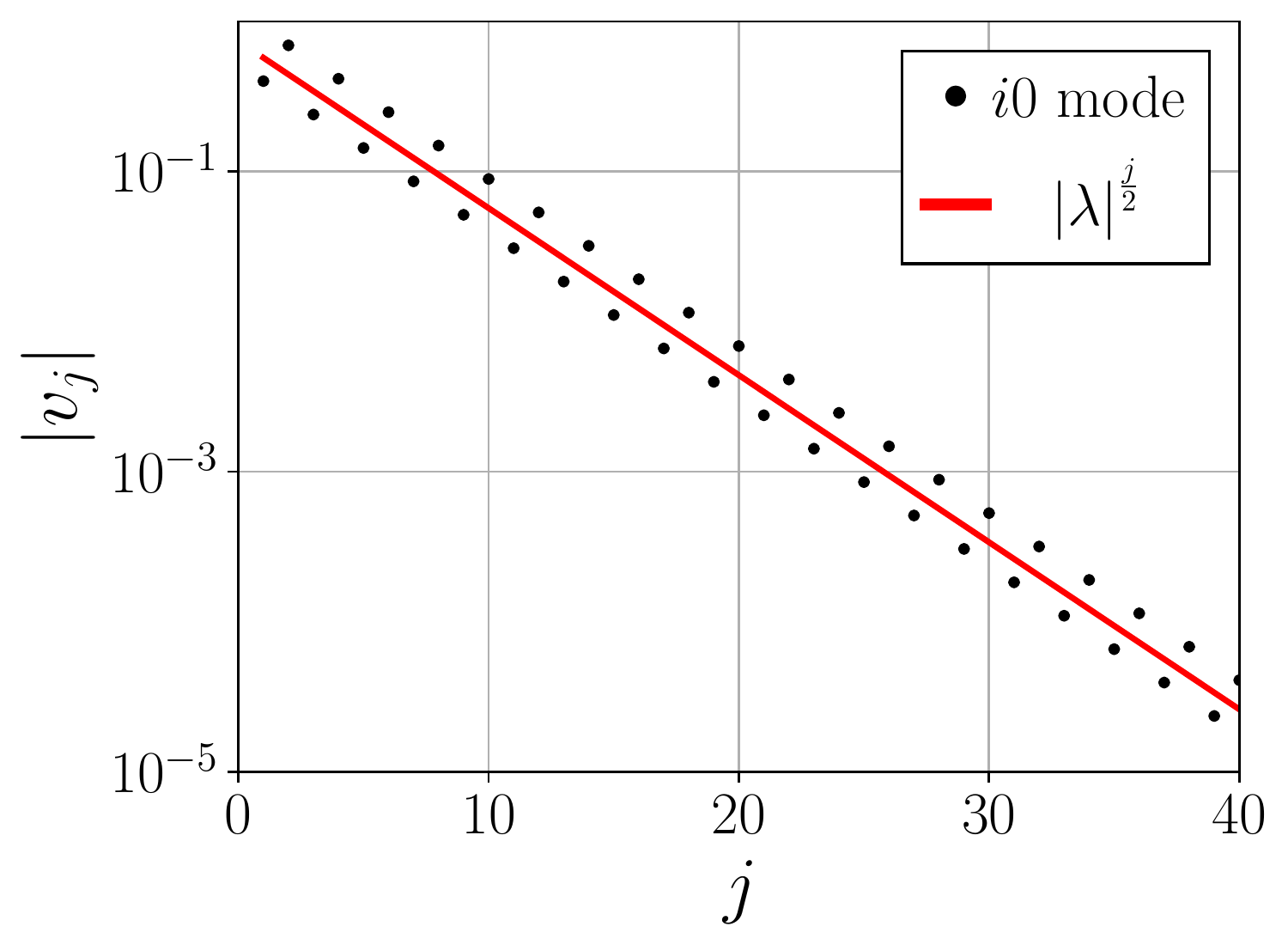}
    \caption{A plot showing the exponential decay of the $i0$ mode for $L=1000, h_y = \frac{\pi}{3}$ and $\beta=2.0$, compared against the decay rate obtained analytically from \cref{eq:slope}.}
    \label{fig:emodefit}
\end{figure}

In the parameter regime shown in \cref{fig:spect1} with $h_y<\frac{\pi}{4}$, there are $i0$ modes as well, except with equal real parts. In this regime, although there are doubly degenerate steady states, oscillations are absent (i.e. $\theta = 0$). These $i0$ modes are obtained by replacing (2) with $[\hat{F}, \hat{V}]=0$. 

For $\hat{F}$, $\vec{v}$ can be computed in the thermodynamic limit by using a transfer matrix method, which proceeds by rewriting the equation $\acomm{\hat{F}}{\hat{V}}=0$ as
\begin{equation}
    \mqty(v_{2j+2}\\v_{2j+1}) = T \mqty(v_{2j}\\v_{2j-1})
\end{equation}
for $1\leq j\leq L-2$. Crucially, this equation only holds for the bulk. We can choose to fulfill either the boundary equation which relates $v_2$ to $v_1$ (to obtain $\hat{F}_L$), or $v_{2L}$ to $v_{2L-1}$ (in the case of $\hat{F}_R$).

Here, we analytically solve for $\vec{v}$ for $J_{xx} = J_{zz} = 0$. By imposing the boundary conditions for $v_1 $ and $v_2$, we find that
\begin{equation}
\label{eq:slope}
    \begin{aligned}
        \mqty(v_{2j}\\v_{2j-1}) = \lambda_1^{j-1}\mqty(\cos(h_y)\\ \sin(h_y))\\
        \lambda_1 \equiv i\cot(h_y)\coth(\beta)
    \end{aligned}
\end{equation}
Requiring that this edge mode decays exponentially fast into the bulk, we have the condition
 \begin{equation}
    |\lambda_1|<1 \implies \cosh(2\beta)\cos(2h_y) < -1, 
\end{equation}
which is \textit{exactly} the condition for the band gap closing in the imaginary direction obtained in \cref{eq:crit}. Thus, we have demonstrated a non-Hermitian bulk boundary correspondence. Extensions of Hermitian topological invariants have previously been used in  studies that numerically obtained non-Hermitian edge modes \cite{LZhou2018,LZhou2020}, adding to this correspondence.
 
We further compare this with the numerical results shown in \cref{fig:emodefit}. For $J_{xx}, J_{zz}\neq 0$, we can also numerically demonstrate that the $i0$ modes are localized on the edges.

Lastly, we introduce random spatial inhomogeneity in the $Y-$ fields. $h_y$ in \cref{eq:vdefine} now assumes a position index $j$, i.e. $h_{y,j} = \frac{\pi}{3} \pm \delta\widetilde{h}$, where $\widetilde{h}$ is uniformly drawn from $\qty[-1,1]$ for each site. Even in the presence of strong disorder, the $i0$ modes still persist \cite{suppMat}, provided that the gap does not close in the imaginary direction.

The robustness of the edge modes requires $L\to\infty$ since the two edge modes can couple in small systems to produce a splitting in both the real and imaginary directions, resulting in a single steady state. The analysis of the splitting with finite $L$ is presented in \cite{suppMat}.

\begin{figure}
    \hspace{-2em}\includegraphics[width=0.5\textwidth]{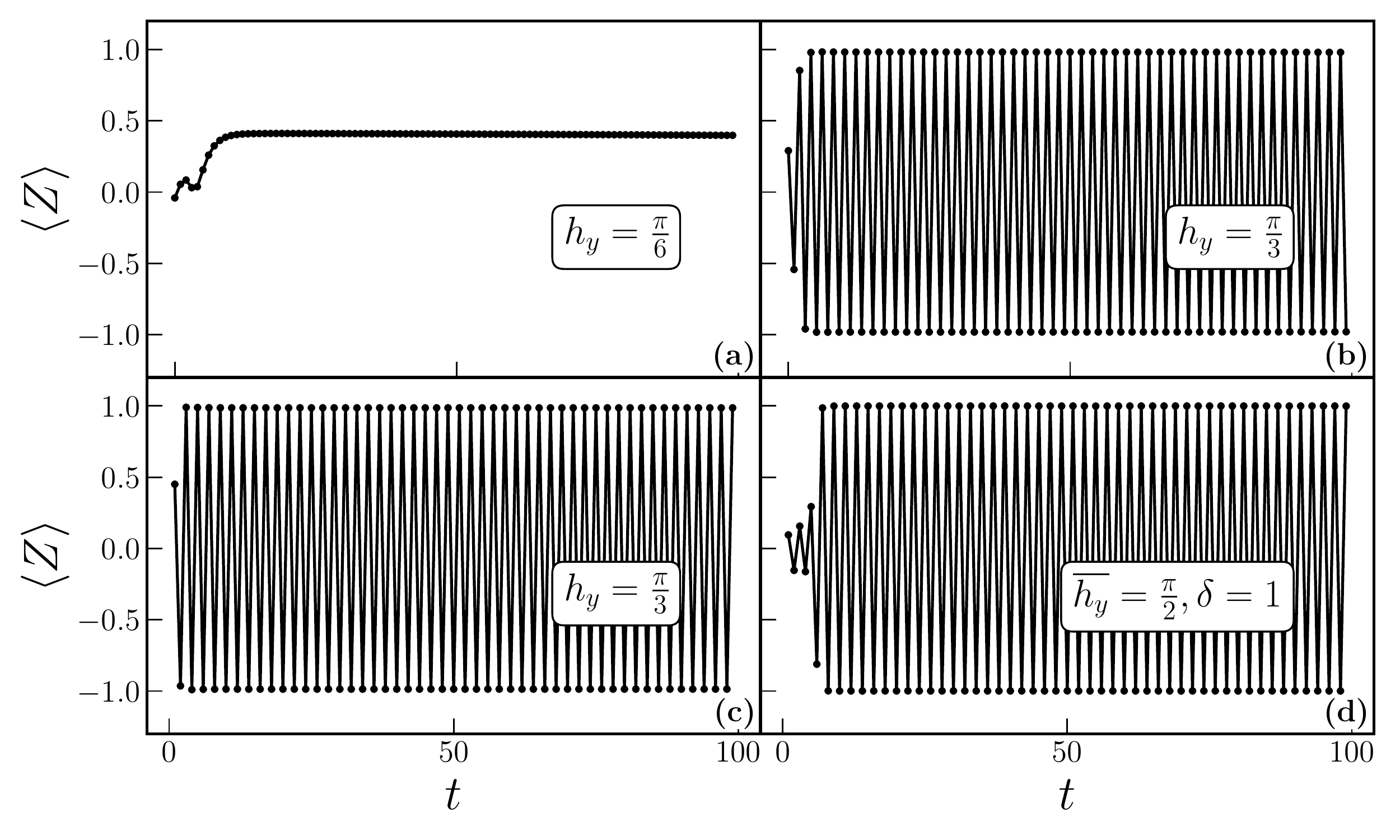}
    \caption{Plots of $\expval{Z}$ without (a,b) and with interactions ((c,d), $J_{yy} = 0.3$). (a) Oscillations are absent and the $i0$ modes are degenerate. (b) Oscillations with double the period are present, and the $i0$ modes show a $\pi$ splitting. (c) A clean interacting system. (d) A system with strong stochasticity in $h_{y,j}$. In all cases, $J_{xx}=0.3, \beta=0.75, L=100$.}
    \label{fig:dynTrans}
\end{figure}

\textit{Dynamics and Interaction} -- We now turn to the dynamics to study the signature of these $i0$ modes. The dynamics can be simulated in two ways - first, by exploiting the mapping to free fermions and using the machinery of Fermionic Gaussian States (FGS)\cite{Bravyi2005,PhysRevA.65.032325,Bravyi2012,10.21468/SciPostPhysLectNotes.54,10.21468/SciPostPhys.7.2.024,PhysRevB.103.224210}, and second, using MPS methods \cite{PAECKEL2019167998,PhysRevLett.93.076401} in terms of the spin degrees of freedom.

A limitation of FGS is that only Fermionic states with a definite parity can be simulated. Since $\hat{V}$ conserves parity, the oscillations cannot be observed directly using FGS, since the two steady states $\ket{\psi_\pm} \sim \ket{\uparrow\uparrow\uparrow\cdots} \pm i^L\ket{\downarrow\downarrow\downarrow\cdots}$ have different parities. Instead, beginning with states of different parities, one can show that the final states at long times have an overlap $\sim 1$ with one of $\ket{\psi_\pm}$, providing indirect evidence for the presence of oscillations. A second limitation is that FGS cannot describe models with interactions. Thus, we use MPS methods to study the dynamics, utilizing the ITensor C++ Package \cite{itensor}.

We consider random product initial states $\ket{\psi}_0 = \ket{\uparrow\uparrow\downarrow\uparrow\downarrow\cdots}$, where each spin points up or down along the z-axis. This state is stroboscopically evolved using \cref{eq:erule}, and the quantity $\expval{Z(t)}\equiv \frac{1}{L}\sum\limits_j \ev{\hat{Z}_j}{\psi(t)}$ is calculated. Oscillations in $\ev{Z}$ occur when there are $i0$ modes with a difference of $\pi$ in the real parts of their quasi-energies, and not otherwise (See \cref{fig:dynTrans}).

Interactions are introduced including $e^{-iJ_{yy}\sum\limits_j \hat{Y}_j \hat{Y}_{j+1}}$ in $\hat{U}_R$, which corresponds to a 4-fermion interaction $e^{-iJ_{yy} \sum\limits_j \gamma_{2j-1}\gamma_{2j}\gamma_{2j+1}\gamma_{2j+2}}$. Such interactions can lead to thermalization in unitary models, which usually destabilizes any order \cite{PhysRevB.93.104203,PhysRevE.90.012110}. However, as shown in Fig.~\ref{fig:dynTrans}, oscillations persist in the presence of interactions.

 Finally, we consider $Y$-fields that are random in both time and space, modeled as $e^{-i\sum\limits_j h_{y,j}(t) \hat{Y}_j}$, where $h_{y,j}(t) = \overline{h_y} + \delta\widetilde{h}_{y,j}(t)$, and $\widetilde{h}_{y,j(t)}$ is drawn uniformly from $\qty[-1,1]$ at every time step. Again, oscillations persist, both in the interacting and the non-interacting models, confirming the stability of the $i0$ mode.

\begin{figure}
    \hspace{-1em}\includegraphics[width=0.5\textwidth]{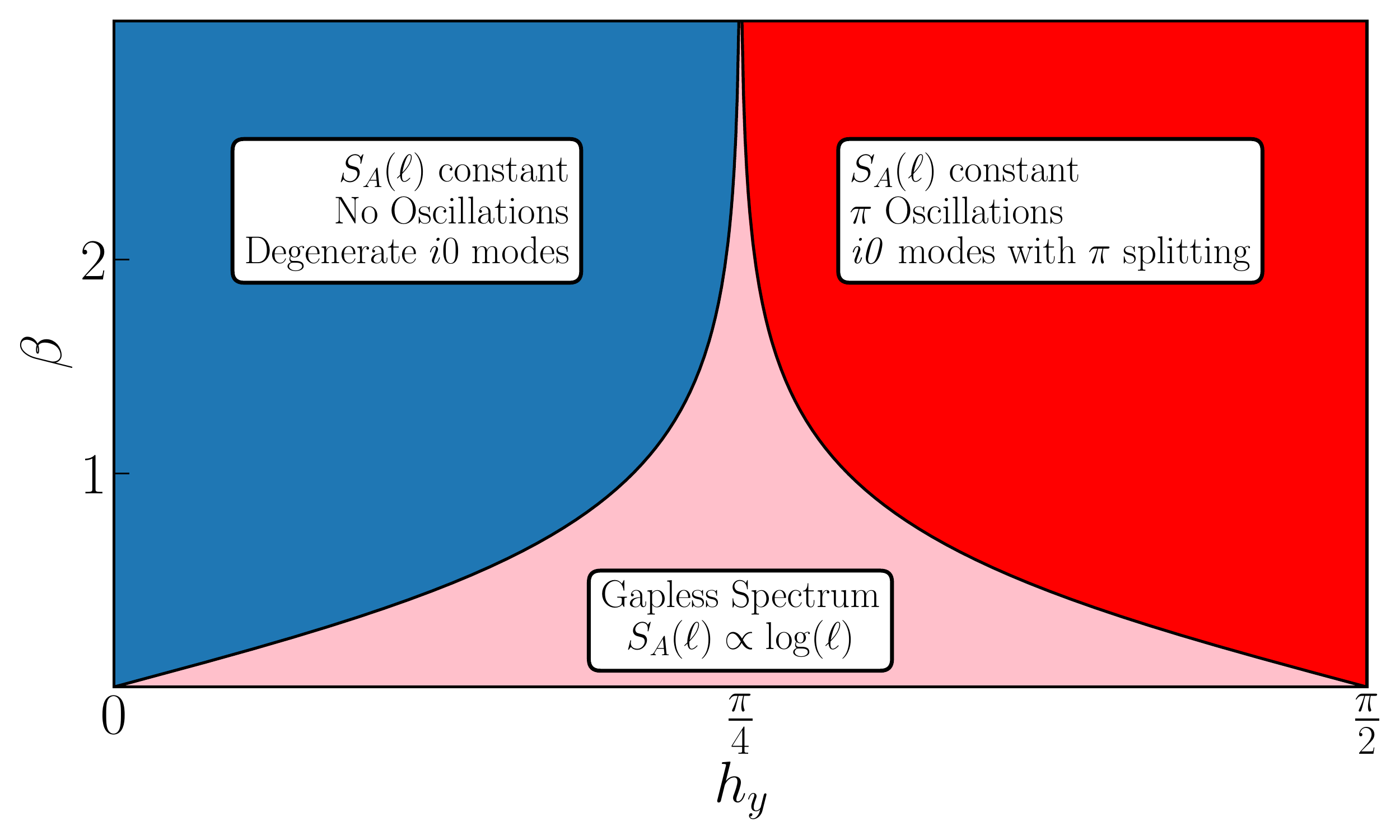}
    \caption{A phase diagram summarizing the three different phases that are observed as $\beta$ and $h_y$ are tuned, in the case where $J_{yy}=J_{zz}=J_{xx} = 0$ and the phase boundary is analytically determined. The phase diagram remains qualitatively the same for nonzero $J_{xx}$ and $J_{zz}$, with possibly more gap closings for larger $J_{xx}, J_{zz}$. The gapless critical phase is a special feature of non-unitary free fermion dynamics and will be replaced by a volume law phase in the presence of the interaction, e.g., $J_{yy}\neq 0$.}
    \label{fig:PhaseDiag}
\end{figure}

\textit{Discussion} -- In this work, we have studied the emergence of oscillatory behaviour in a periodically driven nonunitary system of qubits. We have found that a critical strength of measurement is required to observe oscillations that break the discrete time-translation symmetry in these systems. Such dynamical behavior is accompanied by the emergence of a non-Hermitian $i0$ mode, which is robust to various perturbations, both quenched and stochastic. Such models can be realized in quantum circuits where 1- and 2-site unitary gates as in $\hat{U}_R$ are applied to the qubits. The imaginary time evolution can be implemented by subjecting the system to weak measurements corresponding to the following Kraus operators at each site $j$

\begin{equation}
\begin{aligned}
    M^\pm_j = \frac{1}{\sqrt{2\cosh\qty(2\beta)}}\qty(\cosh(\beta) \pm \sinh(\beta) \hat{Z}_j \hat{Z}_{j+1})
\end{aligned}
\end{equation}

and post-selecting for the $+$ outcome.

The steady states in either phase studied in the text obey an area law entanglement scaling. However, there is an intermediate regime between the two steady states where the spectrum is gapless in the imaginary direction. The transition from a regime where there is an imaginary gap in the spectrum of $\hat{H}_F$, to one where there isn't, is reflected in a change in the entanglement behavior of the steady states, from an area law to a (parameter-dependent) critical phase \cite{Chen_2020,Alberton_2021,jian2020criticality}. The properties of this phase are detailed in \cite{suppMat}. \cref{fig:PhaseDiag} shows the phase diagram for these non-interacting models as both $h_y$ and $\beta$ are varied.

Whereas we find oscillations even in clean systems, traditional time crystals rely on strong disorder to evade thermalization and thus exhibit order. In our models, the periodic weak measurements may be interpreted as effectively ``cooling" the system to the steady-state subspace. The localization of the $i0$ mode on the edge and its spectral gap to bulk states lend additional explanations for this stability \cite{asboth2016short,Kitaev_2001}, while unitary time crystals do not rely on such edge modes\cite{PhysRevLett.116.250401}.

In future work, we hope to characterize the $i0$ mode in the presence of interactions. It is possible that such an operator might only pair states in a part of the spectrum of $\hat{H_F}$, but still provide detectable signatures in the dynamics. We would also like to study the role of symmetry breaking in these oscillations, and especially, how it might be used to generate oscillations of period greater than 2, and for the case of $N_S > 2$ steady states. Lastly, it would be interesting to understand if analogous $i0$ modes can be found in models that involve a rectification of the system based on measurement outcomes \cite{PhysRevLett.129.090404}.

{\it Acknowledgements.---}
We gratefully acknowledge computing resources from Research Services at Boston College and the assistance provided by Wei Qiu. This research
is supported in part by the National Science Foundation under Grant No. DMR-2219735.

\nocite{*}
\bibliography{biblio}

\end{document}


\title{Supplemental Material:\\Robust Oscillations and Edge Modes in Nonunitary Floquet Systems }
\author{Vikram Ravindranath}
\author{Xiao Chen}
\affiliation{Department of Physics, Boston College, Chestnut Hill, MA 02467, USA}

\onecolumngrid
\setcounter{equation}{0}
\setcounter{figure}{0}
\setcounter{table}{0}
\setcounter{page}{1}
\makeatletter
\renewcommand{\theequation}{S\arabic{equation}}
\renewcommand{\thefigure}{S\arabic{figure}}
\renewcommand{\bibnumfmt}[1]{[S#1]}
\renewcommand{\citenumfont}[1]{S#1}

\maketitle

\section{Diagonalization of non-Hermitian Free Fermion Hamiltonians}
\subsection{Jordan-Wigner Transformation}

The Jordan-Wigner transformation is a tool that is extensively used to map fermionic to spin-$\frac{1}{2}$ degrees of freedom. When $J_{yy} = 0$ in $\hat{V}$, this family of spin models maps to models of free fermions under the Jordan-Wigner transformation. The single-particle spectrum of these non-interacting models is easy to obtain numerically, even for very large system sizes. The full many-body spectrum is then built up by filling in the single particle eigenstates of $\hat{V}$. This transformation proceeds by identifying the spin operators with non-local fermionic operators. Nonlocality is required to accommodate the commutativity of spin operators on different sites. This transformation is defined using the prescription

\begin{equation}
    \begin{aligned}
    	\hat{Y}_j &\to i \gamma_{2j-1} \gamma_{2j}\\
    	\hat{X}_j &\to \left(\prod_{l<j} i \gamma_{2l-1} \gamma_{2l}\right) \gamma_{2j-1} \\
    	\hat{Z}_j &\to \left(\prod_{l<j} i \gamma_{2l-1} \gamma_{2l}\right) \gamma_{2j}
    \end{aligned}
\end{equation}
with $\left\lbrace\gamma_j\right\rbrace_{j=1}^{2L}$ being Majorana operators that obey $\acomm{\gamma_k}{\gamma_l} = 2\delta_{kl}$. For ease of notation, we define $a_j = \gamma_{2j}$ and $b_j = \gamma_{2j-1}$, following \cite{PRXQuantum.2.040319}. Under this transformation, the $XX$ and $ZZ$ couplings are expressed as

	\begin{equation}
	\begin{aligned}
	\hat{Z}_j \hat{Z}_{j+1} &\to -i b_{j} a_{j+1}.
	\end{aligned}
	\end{equation}

By defining a column vector $\vec{\gamma}$ whose entries are the $2L$ majorana operators $\qty{\gamma_j}$
\begin{equation}
\label{seq:vdefine}
    \vec{\gamma} = \mqty(b_1\\ a_1 \\ \\ \vdots\\ \\ \\b_L\\a_L),
\end{equation}
any noninteracting, fermionic, Hermitian Hamiltonian can be written in the form 
\begin{equation}
    H = \frac{\gamma^T \mathcal{H} \gamma}{4} = \frac{1}{4} \sum\limits_{i,j} \gamma_i \mathcal{H}_{i,j} \gamma_j,
\end{equation}
where $\mathcal{H}$ is a $2L\cross2L$ purely imaginary antisymmetric matrix. For instance, $\mathcal{H}_{xx}$ is a tridiagonal matrix, composed of blocks of the Pauli matrix $Y$

\begin{equation}
    \mathcal{H}_{xx} = \mqty(0\\&0 & i\\&-i&0\\&&& \ddots \\&&&&0 & i\\&&&&-i&0\\&&&&&& 0).
\end{equation}
We relax the restriction that the entries of $\mathcal{H}$ are purely imaginary when we consider non-Hermitian Hamiltonians.

\subsection{Spectrum of Free Fermion Hamiltonians}

Given a non-interacting (not necessarily Hermitian) Hamiltonian $\hat{H}$, we now discuss the steps involved in obtaining its spectrum. We start by reviewing this process in the case where $\hat{H}$ is Hermitian.

The most general Hermitian, parity-conserving, quadratic, fermionic Hamiltonian can be written in terms of Majorana operators $\qty{\gamma_j}$ as 
	\begin{equation}
	    \label{eq:quadham}
	    H = \frac{\gamma_i \mathcal{H}_{ij} \gamma_j}{4},
	\end{equation}
	
	where $\mathcal{H}$ can be written as $i$ times a real, $2L\cross2L$ antisymmetric matrix $\mathcal{G}$, such that
	
	\begin{equation}
	    \mathcal{H}^\dagger = \qty(i\mathcal{G})^\dagger=-i\mathcal{G}^T = \mathcal{H}.
	\end{equation}
	
	The spectrum of real even-dimensional antisymmetric matrices comes in pairs of $\pm i\lambda_j; \lambda_j\in\mathbb{R}$, with corresponding eigenvectors $v_j, v_j^*$, where the elements of $v_j^*$ are the complex conjugates of those of $v_j$. $\mathcal{G}$ then has the decomposition
	
	\begin{equation}
	    \label{eq:decomp1}
	    X^T \mathcal{G} X = \mqty(\dmat{\admat[0]{\lambda_1,-\lambda_1},\admat[0]{\lambda_2,-\lambda_2},\dmat{\ddots},\admat[0]{\lambda_L,-\lambda_L}}) \equiv \Sigma,
	\end{equation}
	
	with $X$ a real orthogonal matrix
	
	\begin{equation}
	    X^TX = X X^T = \mathbb{1};\text{ }X_{ij} \in \mathbb{R}.
	\end{equation}.

    The matrix $X$ is constructed from the normalized eigenvectors of $\mathcal{G}$ as
	
	\begin{equation}
	    \label{eq:Xdef}
	    X = \frac{1}{\sqrt{2}} \mqty(\vert&\vert&&\vert&\vert\\v_1 + v_1^*&i(v_1 - v_1^*)&\cdots&v_L + v_L^* & i(v_L - v_L^*)\\ \vert&\vert&&\vert&\vert)
	\end{equation}
	
	If we define a new set of majorana operators $\qty{g_j}$
	\begin{equation}
	    \label{eq:gdef}
	    \vec{g} = X^T \vec{\gamma}
	\end{equation}
	
	which also obey canonical anticommutation relations
	\begin{equation}
	    \label{eq:canon1}
	    \begin{aligned}
	    \acomm{g_i}{g_j} &= \sum\limits_{l,m} X^T_{im} X^T_{jl} \acomm{\gamma_m}{\gamma_l}\\
	    & = 2 \sum\limits_{l,m} X^T_{im} X^T_{jl} \delta_{lm}\\
	    & = 2 (X^T X)_{ij} = 2\delta_{ij},
	    \end{aligned}
	\end{equation}
	
	this decomposition allows us to rewrite the Hamiltonian from \cref{eq:quadham} as
	
	\begin{equation}
	    \label{eq:diagmajham}
	    H = \frac{1}{2}\sum\limits_{j=1}^L i\lambda_j g_{2j-1} g_{2j}.
	\end{equation}
    
    Lastly, defining a set of complex fermionic operators $\qty{f_j, f^\dagger_j}$
    \begin{equation}
        \label{eq:fdef1}
        f_j = \frac{g_{2j} - i g_{2j-1}}{2},
    \end{equation}
    
    we have diagonalized $H$
    \begin{equation}
        \label{eq:diagfermham}
        H = \sum\limits_j \lambda_j \qty(f^\dagger_j f_j - \frac{1}{2}).
    \end{equation}
    
    The many-body eigenstates of $H$ can then be constructed by filling in the single particle states $f^\dagger_j$.
    \begin{equation}
        \label{eq:mbspect1}
        \begin{aligned}
            \ket{n_1,n_2,\cdots,n_L} &= \qty(f^\dagger_1)^{n_1}\qty(f^\dagger_2)^{n_2}\cdots\qty(f^\dagger_L)^{n_L}\ket{0}\\
            H\ket{n_1,n_2,\cdots,n_L} &= (E_0 + \sum\limits_j \lambda_j n_j) \ket{n_1,n_2,\cdots,n_L}
        \end{aligned}
    \end{equation}
    
    Turning to non-Hermitian Hamiltonians, we no longer consider an $\mathcal{H}$ which has purely imaginary entries. However, owing to the anticommutativity of $\qty{\gamma_j}$, $\mathcal{H}$ can still be expressed as an antisymmetric matrix. Further, one now has to distinguish between the right and left eigenvectors (both of $H$ and $\mathcal{H}$), which are not simply related by Hermitian conjugation, as in the Hermitian case. Once these caveats are accounted for, diagonalization proceeds in analogous fashion. The following text expands and elaborates on the methods introduced in \cite{Guo_2021}.
    
    The eigenvalues of a complex, antisymmetric matrix $\mathcal{H}$ can still be written as pairs of $\pm\lambda_j$, with $\lambda_j\in\mathbb{C}$, now. Their corresponding (right) eigenvectors are no longer related by complex conjugation. Therefore, we update our notation as follows. We assume an unambiguous ordering of $\pm\lambda_j$ in the pair $\qty(\lambda_j,-\lambda_j)$. This can be achieved, for instance, by choosing $\lambda_j$ to have a positive real part, or a positive imaginary part, if $\lambda_j\in i\mathbb{R}$. We use this to label the right eigenvectors as
    
    \begin{equation}
        \label{eq:REVecNotn}
        \begin{aligned}
            \mathcal{H}v_{2j-1} &= \lambda_j v_{2j-1}\\
            \mathcal{H}v_{2j} &= -\lambda_j v_{2j}
        \end{aligned}
    \end{equation}
    
    The corresponding $2L\cross1$ dimensional left eigenvectors are labelled 
    \begin{equation}
        \label{eq:LEVecNotn}
        \begin{aligned}
            u_{2j-1}\mathcal{H} &= \lambda_j u_{2j-1}\\
            u_{2j}\mathcal{H} &= -\lambda_j u_{2j}
        \end{aligned}
    \end{equation}
    
    The eigenvectors of an antisymmetric matrix $\mathcal{H}$ have the following properties :
    \begin{enumerate}
        \item If $v$ is a right eigenvector with eigenvalue $\lambda$, $v^T$ is a left eigenvector with eigenvalue $-\lambda$.
        \[ \mathcal{H} v = \lambda v \implies \qty(\mathcal{H} v)^T = \lambda v^T\implies v^T\mathcal{H} = -\lambda v^T\]
        
        \item With the ordering prescription described above, we have the following inner product rules
        \begin{equation*}
            \begin{aligned}
                    &\text{i) } v^T_{2j-1} v_{2k} \propto \delta_{j,k}\\
                    &\text{ii) } v^T_{2j} v_{2k-1} \propto \delta_{j,k}\\
                    &\text{iii) } v^T_{2j-1}v_{2k-1} = 0 = v^T_{2j} v_{2k} 
            \end{aligned}
        \end{equation*}
        
        \begin{equation}
            \begin{aligned}
                &\text{i) } v^T_{2j-1}\mathcal{H }v_{2k} = -\lambda_k v^T_{2j-1} v_{2k} = -\lambda_j v^T_{2j-1} v_{2k}\\
                &\implies \qty(\lambda_j-\lambda_k) v^T_{2j-1} v_{2k} = 0 \\
                &\text{\textbf{Case 1: }}k\neq j\implies \lambda_j \neq \lambda_k \implies v^T_{2j-1} v_{2k} \propto \delta_{j,k}\\
                &\text{\textbf{Case 2: }}k\neq j \text{ but } \lambda_k = \lambda_j = \lambda\\
                &\text{\hspace{4em}}\text{We have }\\
                &\text{\hspace{4em}}\text{\hspace{1em}}\mathcal{H} \qty(v_{2j},v_{2k}) = -\lambda\\
                &\text{\hspace{4em}}\text{If } v^T_{2j-1} v_{2k} \neq 0 \text{ and } v^T_{2j-1} v_{2j} \neq 0, \text{we can redefine }\\
                &\text{\hspace{4em}}\text{\hspace{1em}}v_{2k} \to v_{2k} - \frac{v^T_{2j-1} v_{2k}}{v^T_{2j-1} v_{2j}} v_{2j}\implies v^T_{2j-1} v_{2k}.\\
                &\text{\hspace{4em}}\text{Lastly, if } v^T_{2j-1} v_{2j} = 0, \text{ but } v^T_{2j-1} v_{2k} \neq 0, \text{ we can simply swap}\\
                &\text{\hspace{4em}}v_{2k} \leftrightarrow v_{2j}, \text{ thus } v^T_{2j-1} v_{2k} \propto \delta_{j,k}
            \end{aligned}
        \end{equation}
        
        ii) follows from transposing i).
        
        \begin{equation}
            \begin{aligned}
                &\text{iii) } v^T_{2j-1}\mathcal{H }v_{2k-1} = \lambda_k v^T_{2j-1} v_{2k} = -\lambda_j v^T_{2j-1} v_{2k}\\
                &\implies \qty(\lambda_j+\lambda_k) v^T_{2j-1} v_{2k-1} = 0 \implies v^T_{2j-1} v_{2k-1} = 0\\
                &v^T_{2j}\mathcal{H }v_{2k} = -\lambda_k v^T_{2j-1} v_{2k} = \lambda_j v^T_{2j-1} v_{2k}\\
                &\implies\qty(\lambda_j+\lambda_k) v^T_{2j} v_{2k} = 0 \implies v^T_{2j} v_{2k} = 0\\
            \end{aligned}
        \end{equation}
        
        The assertion that $\lambda_j + \lambda_k \neq 0$ can be made because the eigenvalues and their corresponding eigenvectors have been ordered according to a particular rule that ensures that a pair of eigenvalues $\pm\lambda$ is always ordered in the same way, regardless of the position of their occurence in the spectrum.
        
        \item With the normalization that $v^T_{2j-1} v_{2j} = v_{2j}^Tv_{2j-1} = 1$, the left eigenvectors corresponding to the eigenvalue pair $\pm\lambda_j$ are $\qty(u_{2j-1,}u_{2j}) \equiv \qty(v^T_{2j,}v^T_{2j-1})$. Moreover, we have $u_j v_k = \delta_{j,k}$.
    \end{enumerate}
    
    We are now ready to construct an analogous $X$ for the general antisymmetric matrix, defined as
    
    \begin{equation}
        \label{eq:Xdef2}
        \begin{aligned}
        X &= \frac{1}{\sqrt{2}} \mqty(\vert&\vert&&\vert&\vert\\v_1 + v_2&i(v_1 - v_2)&\cdots&v_{2L-1} + v_{2L} & i(v_{2L-1} - v_{2L})\\ \vert&\vert&&\vert&\vert)\\
            &X^T X = X X^T = \mathbb{1}
        \end{aligned}
    \end{equation}
    
    It can be straightforwardly verified that 
    
    \begin{equation}
	    \label{eq:decomp2}
	    X^T \mathcal{H} X = \mqty(\dmat{\admat[0]{i\lambda_1,-i\lambda_1},\admat[0]{i\lambda_2,-i\lambda_2},\dmat{\ddots},\admat[0]{i\lambda_L,-i\lambda_L}}) \equiv \Sigma,
	\end{equation}
    
    since
    
    \begin{equation}
            \mathcal{H} X = \frac{1}{\sqrt{2}} \mqty(\vert&\vert&&\vert&\vert\\\lambda_1(v_1 - v_2)&i\lambda_1(v_1 + v_2)&\cdots&\lambda_L(v_{2L-1} - v_{2L}) & i\lambda_L(v_{2L-1} + v_{2L})\\ \vert&\vert&&\vert&\vert)
    \end{equation}
    
    and
    
    \begin{equation}
        \begin{aligned}
            &\frac{1}{2} (v_{2k-1} + v_{2k})^T (v_{2j-1} + v_{2j}) = -\frac{1}{2} (v_{2k-1} - v_{2k})^T (v_{2j-1} - v_{2j}) = \delta_{k,j}\\
            &\frac{1}{2} (v_{2k-1} + v_{2k})^T (v_{2j-1} - v_{2j}) = -\frac{1}{2} (v_{2k-1} - v_{2k})^T (v_{2j-1} + v_{2j}) = 0
        \end{aligned}
    \end{equation}
    
    We can similarly define a new set of majorana-like operators $\qty{g_j}$ that obey canonical anticommutation relations
    
    \begin{equation}
        \vec{g} = X^T \vec{\gamma}.
    \end{equation}
    
    and the Hamiltonian can be written as
    
    \begin{equation}
        H = \frac{1}{2}\sum\limits_j i\lambda_j g_{2j-1}g_{2j}
    \end{equation}
    
    However, $g_j^\dagger\neq g_j$, since $X$ is a \textit{complex} orthogonal matrix. This leads us to define 2 sets of complex fermionic operators $\qty{f_{L,j}, f^\dagger_{L,j}}$ and $\qty{f_{R,j}, f^\dagger_{R,j}}$, where $L(R)$ denote the left and right eigenstates of the operator $H$, respectively.
    
    \begin{equation}
        \begin{aligned}
            f^\dagger_{R,j} = \frac{g_{2j} + i g_{2j-1}}{2}\\
            f_{L,j} = \frac{g_{2j} - i g_{2j-1}}{2}
        \end{aligned}
    \end{equation}
    
    These operators have the following anticommutation relations
    
    \begin{equation}
        \begin{aligned}
            \acomm{f^\dagger_{R,j}}{f_{L,k}} &= \delta_{j,k}\\
            \acomm{f^\dagger_{R,j}}{f^\dagger_{R,k}} &= 0\\
            \acomm{f_{L,j}}{f_{L,k}} &= 0
        \end{aligned}
    \end{equation}
    
    Crucially, $\qty(f^\dagger_{R,j})^\dagger \neq f_{L,j}$. $H$ can now be expressed in terms of these $f$ operators as
    
    \begin{equation}
        \label{eq:diagfermhamNH}
        \hat{H} = \sum\limits_j \lambda_j \qty(f^\dagger_{R,j} f_{L,j} - \frac{1}{2}).
    \end{equation}
    
    The (right) vacuum state $\ket{0}_R$ of $H$ is defined by
    \begin{equation}
        f_{L,j}\ket{0}_R = 0;\text{ }j=1,2,\cdots,L
    \end{equation}

    The right eigenstates are now constructed from the vacuum state of $H$, by the application of $f^\dagger_{R,j}$.
   \begin{equation}
        \label{eq:mbspect2}
        \begin{aligned}
            \ket{n_1,n_2,\cdots,n_L}_R &= \qty(f^\dagger_{R,1})^{n_1}\qty(f^\dagger_{R,2})^{n_2}\cdots\qty(f^\dagger_{R,L})^{n_L}\ket{0}_R\\
            H\ket{n_1,n_2,\cdots,n_L}_R &= (E_0 + \sum\limits_j \lambda_j n_j) \ket{n_1,n_2,\cdots,n_L}_R
        \end{aligned}
    \end{equation}
    
    We are interested in the right eigenstates of $H$ since these will be used to construct the steady states of our non-unitary time evolution operator. An analogous process can be used to construct the left eigenstates of $H$. 
    
    For example, the steady state under the time evolution given by $e^{-i H}$ is given by the eigenstate whose eigenvalue has the largest imaginary part.
    \begin{equation}
        \label{eq:steadystate}
        \ket{SS}_R = \prod\limits_{\Im{\lambda_j}>0} f^\dagger_{R,j}\ket{0}_R
    \end{equation}

\subsection{Quasi-energy Spectrum of Non-interacting Floquet Operators}

The penultimate step in calculating the spectrum of our non-interacting Floquet Hamiltonian $\hat{H}_F \equiv \frac{\gamma^T \mathcal{H}_F \gamma}{4}$ is to obtain $\mathcal{H}_F$ from $\hat{V}$. This can be done by exploiting the Gaussian nature of the various operators in $\hat{U}_R$ and $\hat{U}_I$. Generally, if one has

\[\hat{V} = e^{\frac{\vec{\gamma}^T A_1\vec{\gamma}}{4}}e^{\frac{\vec{\gamma}^T A_2\vec{\gamma}}{4}}\]

with $A_i^T = -A_i$, then 

\begin{align}
    \hat{V}&= e^{\frac{\vec{\gamma}^T A\vec{\gamma}}{4}};\nonumber\\
    e^{A} &\equiv e^{A_1}e^{A_2}.
    \label{seq:prodrule}
\end{align}

This can be shown by noting that \[{\comm{\frac{\vec{\gamma}^T A_1\vec{\gamma}}{4}}{\frac{\vec{\gamma}^T A_2\vec{\gamma}}{4}}} = \vec{\gamma}^T\frac{\comm{A_1}{A_2}}{4}\vec{\gamma},\] followed by an application of the BCH formula. If $\hat{V}$ is now a Floquet operator, this property allows us to obtain the spectrum of $A$, which we have shown to be the single-particle spectrum of $\frac{1}{4} \gamma^T A \gamma$ (and thus, of $\hat{V}$ as well).

When $\hat{V}$ is invariant under translations, the spectrum can be obtained analytically. We show this for the case where $\hat{V} =  e^{(\beta -i J_{zz})H_{ZZ}} e^{-i J_{xx} H_{XX}} e^{-i h H_Y}$. Explicitly, these Hamiltonians have the following expressions

\begin{equation}
    \begin{aligned}
        H_{XX} = i\sum\limits_{j=1}^L a_j b_{j+1}\equiv \frac{\vec{\gamma}^T\mathcal{H}_{XX}\vec{\gamma}}{4}\\
        H_{ZZ} = -i\sum\limits_{j=1}^L b_j a_{j+1}\equiv\frac{\vec{\gamma}^T\mathcal{H}_{ZZ}\vec{\gamma}}{4}\\
        H_Y = i\sum\limits_{j=1}^L b_j a_{j}\equiv\frac{\vec{\gamma}^T\mathcal{H}_{Y}\vec{\gamma}}{4}
    \end{aligned}
\end{equation}

with $\qty(a_{L+1},b_{L+1}) = \pm (a_1, b_1)$. The choice of (anti-) periodic boundary conditions only constrains the $k$ values to be (half-) integer multiples of $\frac{2\pi}{L}$ and has no effect on the presence of a gap in the spectrum in the thermodynamic limit. We can now write $\hat{V}$ in the form suggested by \cref{seq:prodrule}.

\begin{equation}
    \begin{aligned}
        \hat{V} &= e^{-i\frac{\vec{\gamma}^T\mathcal{H}_F\vec{\gamma}}{4}} \equiv e^{-i \hat{H}_F},\\
        \text{with } e^{-i\mathcal{H}_F} &= e^{(\beta -i J_{zz})\mathcal{H}_{ZZ}} e^{-i J_{xx} \mathcal{H}_{XX}} e^{-i h \mathcal{H}_Y}
    \end{aligned}
\end{equation}

Since each Hamiltonian decomposes into blocks for each $k$, we can write, denoting $\vec{\gamma_k}\equiv \qty(a_k, b_k, a_{-k},b_{-k})$, 
\begin{equation}
    \begin{aligned}
        \hat{H}_F &= \sum\limits_{k>0} \frac{\vec{\gamma_k}^T \mathcal{H}_F(k) \vec{\gamma_k}}{4}\\
        e^{-i\mathcal{H}_F(k)} &= e^{(\beta - iJ_{ZZ})\mathcal{H}_{ZZ}(k)}\\
        &e^{-iJ_{XX}\mathcal{H}_{XX}(k)}e^{-ih\mathcal{H}_{Y}(k)}
    \end{aligned}
\end{equation}

It is useful to note the $k$-space representation of the Hamiltonians
\begin{equation}
    \begin{aligned}
        \mathcal{H}_{XX}(k) &= 2i\mqty(0&\cos(k)&\sin(k)&0\\
        -\cos(k)&0&0&-\sin(k)\\
        -\sin(k)&0&0&\cos(k)\\
        0&\sin(k)&-\cos(k)&0\\)\\
        \mathcal{H}_{ZZ}(k) &= \mathcal{H}_{XX}(-k)\\
        \mathcal{H}_{Y}(k) &= 2\mqty(\dmat{\sigma_y, \sigma_y}) = 2\qty(\mathbb{1}\otimes\sigma_y)
    \end{aligned}
\end{equation}

The energy levels of the Hamiltonian $\hat{H}_F$ correspond to $\frac{1}{2}$ times the eigenvalues of $\mathcal{H}_{F}$. This can be seen by considering each term in \cref{eq:diagfermhamNH}, which can be rewritten as

\begin{equation}
        \hat{H} = \sum\limits_j \frac{\lambda_j}{2} \qty(2f^\dagger_{R,j} f_{L,j} - 1).
\end{equation}

Since the eigenvalues of $\qty(2f^\dagger_{R,j} f_{L,j} - 1)$ are $\pm1$, this results in the contribution of each mode, and thus the single particle spectrum, being $\pm\frac{\lambda_j}{2}$.

In practice, we first find the eigenvalues of $e^{-i\mathcal{H}_F(k)}$ (analytically) or of , calculate their logarithms and then halve them to get the quasi-energy spectrum $\epsilon(k)$ of $\hat{H}_F$. Since $\hat{V}$ is particle-hole symmetric, care must be taken to ensure the symmetry of the spectrum of $\hat{H}_F$ about 0.

As a demonstration, we show how one calculates the spectrum for the case where $J_{XX} = J_{ZZ} = 0$. $\mathcal{H}_{ZZ}(k)$ has an interesting structure, in that it can be written as 
\begin{equation}
    \mathcal{H}_{ZZ}(k)= -2 \qty(\cos(k)\mathbb{1}\otimes\sigma_y + \sin(k)\tau_y\otimes\sigma_z ) 
\end{equation}

where $\tau_{x,y,z}$ and $\sigma_{x,y,z}$ denote the usual Pauli matrices acting in distinct spaces. Going forward, the $\otimes$ will be omitted when its presence is obvious.

An immediate consequence of this structure is that $\qty(\frac{\mathcal{H}_{ZZ}(k)}{2})^2 = \mathbb{1}$, so that
\begin{equation}
\begin{aligned}
    e^{\beta\mathcal{H}_{ZZ}(k)} &= \cosh(2\beta)\mathbb{1} + \sinh(2\beta)\frac{\mathcal{H}_{ZZ}(k)}{2}\\
    &=\cosh(2\beta) - \sinh(2\beta) \qty(\cos(k)\sigma_y + \sin(k)\tau_y\sigma_z ) 
\end{aligned}
\end{equation}

Similarly, 
\begin{equation}
\begin{aligned}
    e^{-ih\mathcal{H}_{Y}(k)} &= \cos(2h) - i\sin(2h)\sigma_y
\end{aligned}
\end{equation}

Multiplying the two matrices, and fixing $\tau_y = \pm 1$ (which leads to a degeneracy), $e^{-i\mathcal{H}_F(k)}$ is of the form $c_0(k) + \vec{c}(k)\dot\vec{\sigma}$. The eigenvalues of $e^{-i\mathcal{H}_F(k)}$ can be obtained straightforwardly as $c_0 \pm \sqrt{\vec{c}.\vec{c}}$. The single particle spectrum of $\hat{H}_F(k)$ is then
\begin{equation}
    \epsilon(k) = \frac{i}{2}\log(c_0 \pm \sqrt{\vec{c}.\vec{c}})
\end{equation}

\begin{figure}
    \centering
    \includegraphics[width=0.475\textwidth]{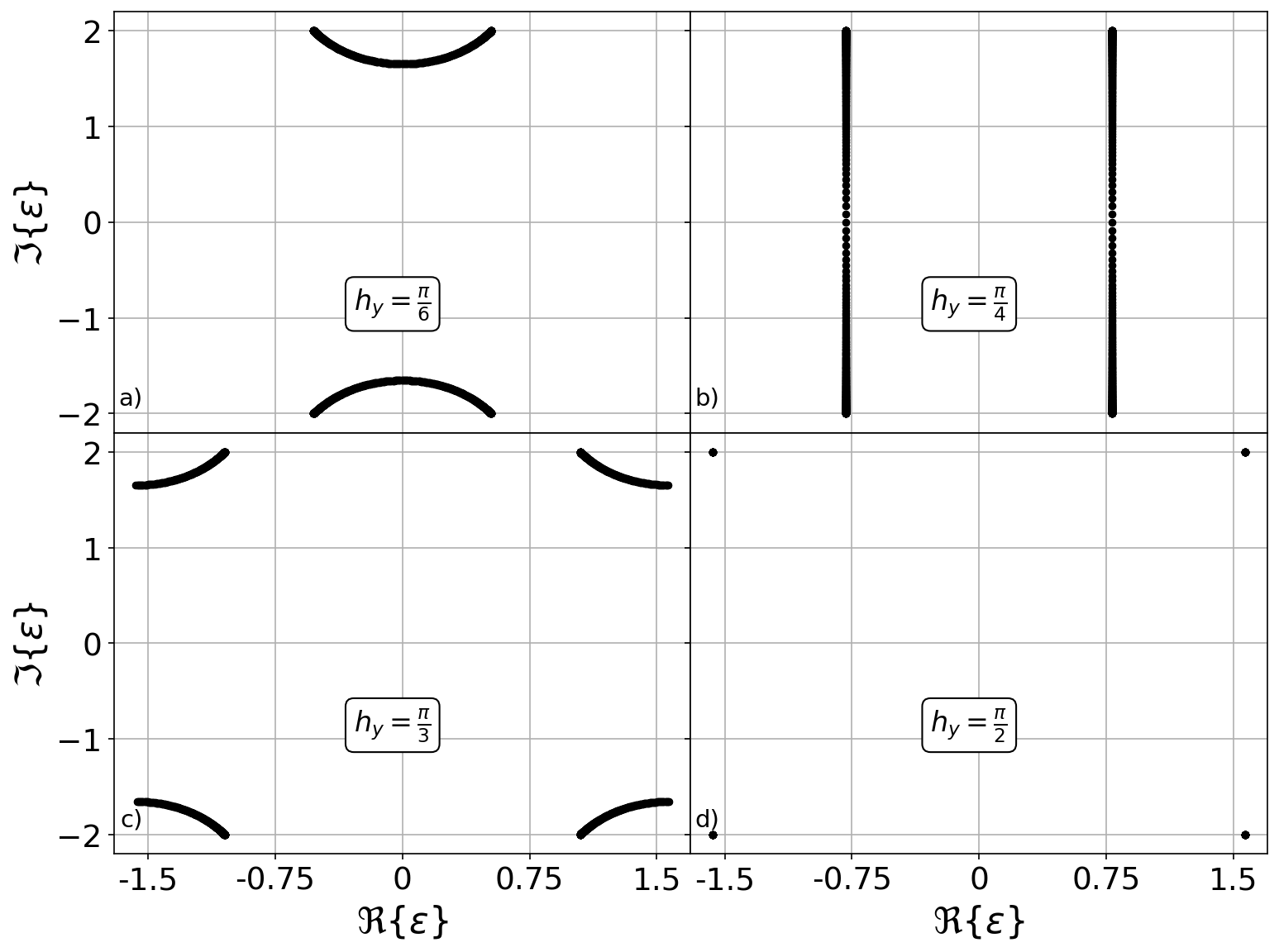}
    \caption{The complex spectrum, showing a gap closing and reopening near $h_y = \frac{\pi}{4}$, as given by Eq. (7) in the main text.}
    \label{fig:spectSimp}
\end{figure}

The numerical diagonalization of $e^{-i\mathcal{H}_F}$ possesses a caveat that is absent in the analytical procedure, owing to the Floquet nature of the problem. A logarithm of each eigenvalue of $\mathcal{H}_F$, followed by its halving, is required to obtain the single particle spectrum of $\hat{H}_F$. As a result, the numerical procedure cannot distinguish between the two $i0$ modes $\epsilon_{i0} = \pm \frac{\pi}{2}$, since these are both reflected as an $e^{\mp i\pi} = -1$ eigenvalue of $e^{-i\mathcal{H}_F}$. Should a point appear only at one of $\frac{\pi}{2}$, an additional verification that it is doubly degenerate is required, and this is indeed the case.

\begin{figure}
    \centering
    \hspace{-2em}
    \includegraphics[width=0.45\textwidth]{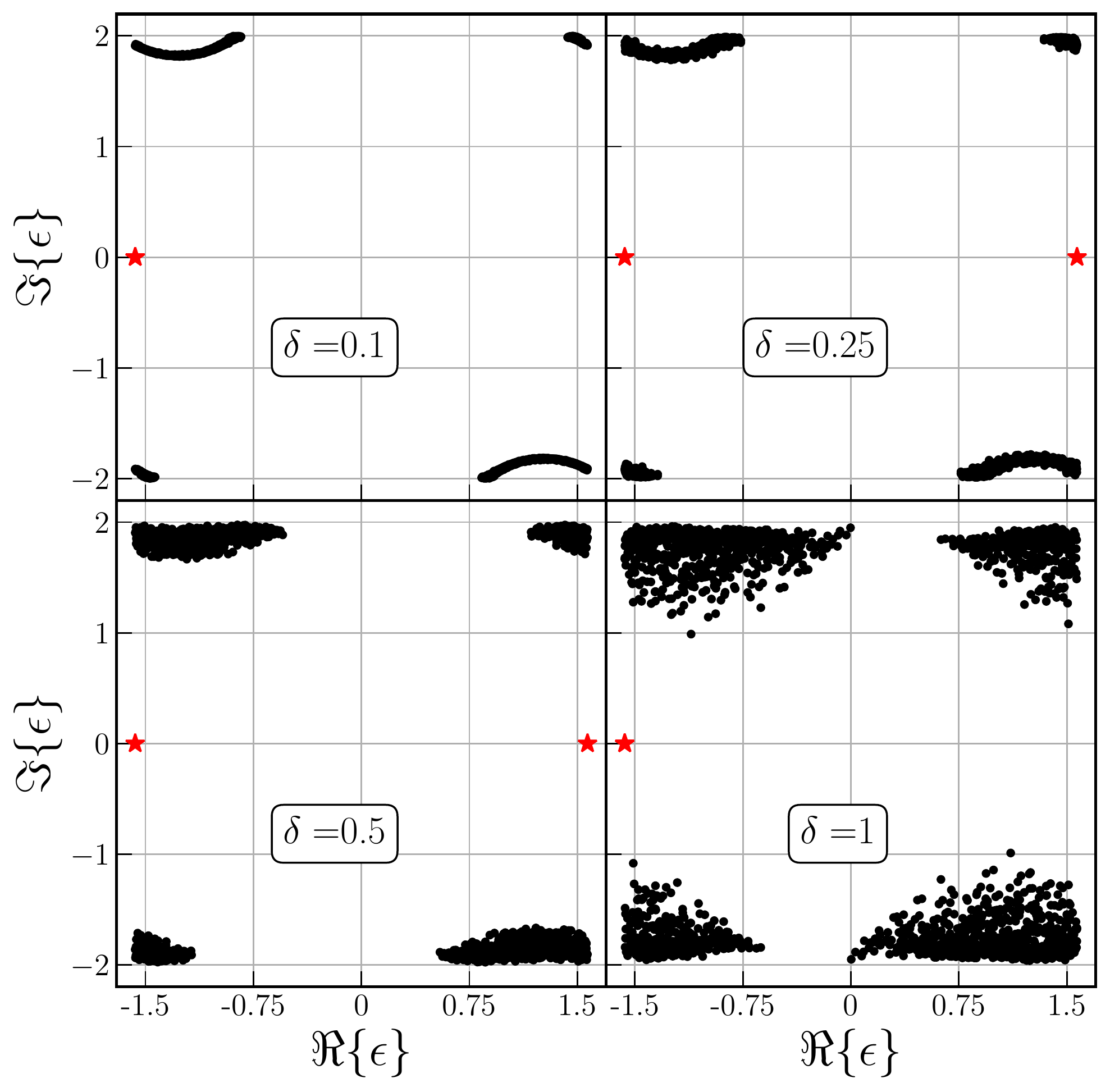}\quad\quad\includegraphics[width=0.45\textwidth]{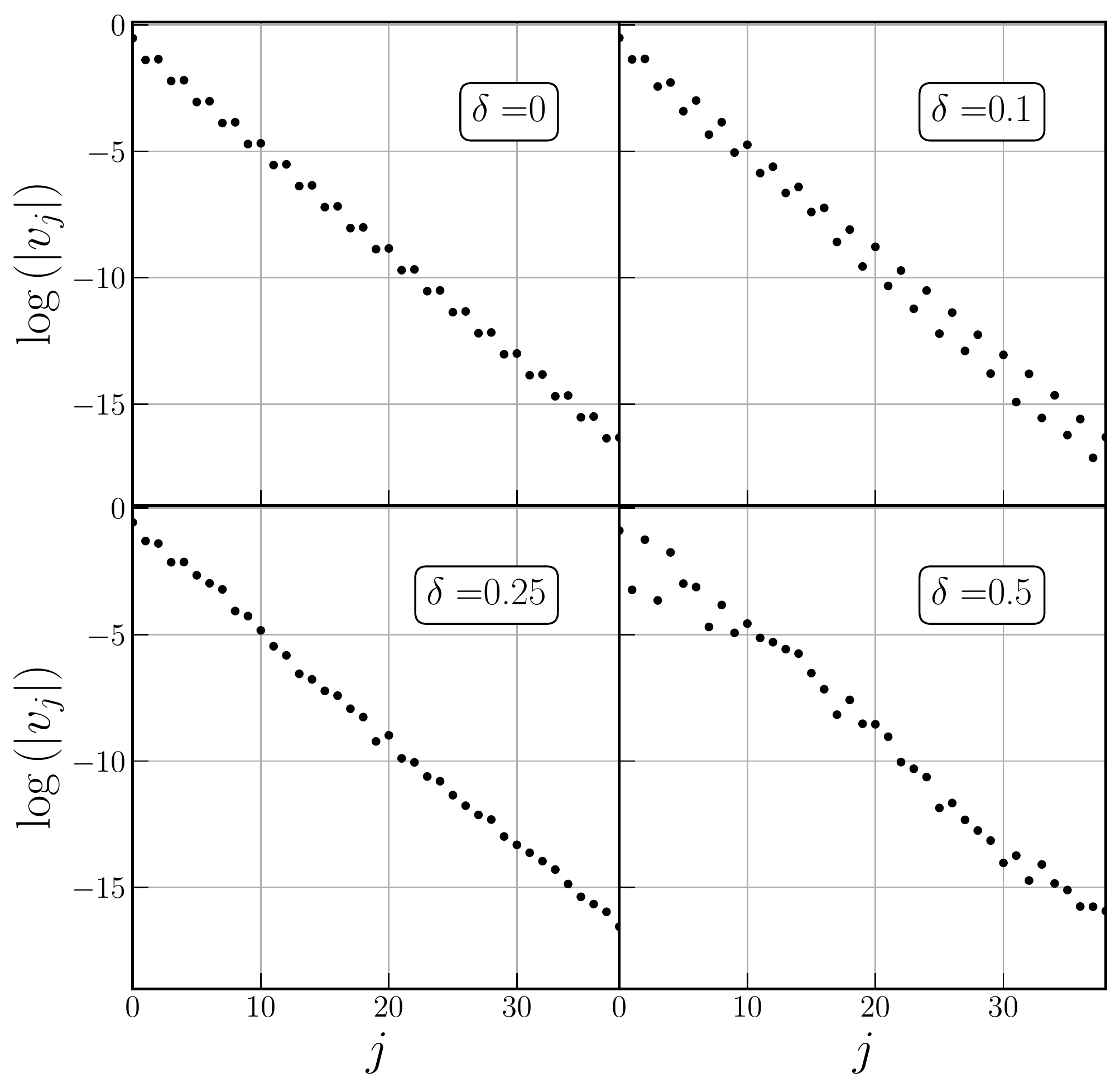}
    \caption{(Left) Spectrum for the disordered case.\ (Right) Corresponding $i0$ mode $\hat{F}_L$.\ Open boundary conditions are considered.}
    \label{fig:spectDis}
\end{figure}

\section{Transfer Matrix Method to obtain Edge Modes}

In this section, we discuss the details of the transfer matrix method used to obtain the edge $i0$ modes. Concretely, we do this for $\hat{V} = e^{\beta \sum Z_j Z_{j+1}} e^{-i \sum hY_{j}}$. In terms of Majoranas, this corresponds to $\hat{V} = e^{-i\beta \sum\limits_{j=1}^{L-1} b_j a_{j+1}} e^{h\sum\limits_{j=1}^L b_j a_j}$. Requiring that $\acomm{\hat{V}}{a_j} = \acomm{\hat{V}}{b_j} = 0$ gives us the $2L$ equations

\begin{align}
    a_1 &: (\alpha_1 + 1) v_1 - \alpha_2 v_2 = 0\\
    \lbrace b_j\rbrace_{j=1}^{L-1} &: \alpha_2 v_{2j-1} + (\alpha_1 + \alpha_3) v_{2j} + \alpha_4 v_{2j+1} = 0\\
    \lbrace a_j\rbrace_{j=2}^{L} &: -\alpha_4 v_{2j-2} + (\alpha_1 + \alpha_3) v_{2j-1} - \alpha_2 v_{2j} = 0\\
    b_L &: (\alpha_1 + 1) v_{2L} + \alpha_2 v_{2L-1} = 0
\end{align}

with $\alpha_1 = \cos{\left(2h\right)} , \alpha_2 = \sin{\left(2h\right)}, \alpha_3 = \cosh{\left(2\beta\right)}$ and $\alpha_4 = i\sinh{\left(2\beta\right)}$, such that $\alpha_1^2 + \alpha_2^2 = \alpha_3^2 + \alpha_4^2 = 1$. Defining a matrix $M$ as 
\begin{equation}
    M = \mqty(\alpha_1 + 1 & -\alpha_2\\
    \alpha_2&\alpha_1 + \alpha_3&\alpha_4\\
    &-\alpha_4&\alpha_1 + \alpha_3&-\alpha_2\\
    &&\ddots&\ddots&\ddots\\
    &&&&\alpha_2&\alpha_1+1),
\end{equation}
the problem now translates to finding the lowest magnitude eigenvectors $\vec{v}$ of $M$, and to check whether their corresponding eigenvalues approach 0 exponentially as $L\to\infty$. We now explicitly construct $\vec{v}$  using a transfer matrix approach.

Presently considering only the "bulk" equations, i.e. those pertaining to $b_j, a_{j+1}$ for $1\leq j\leq L-1$, we can obtain $\mqty(v_{2j+2}\\v_{2j+1})$ in terms of $\mqty(v_{2j}\\v_{2j-1})$ as

\begin{equation}
    \mqty(v_{2j+2} \\ v_{2j+1}) = \underbrace{-\mqty(\frac{\qty(\alpha_1+\alpha_3)^2 + \alpha_4^2 }{\alpha_4\alpha_2}&\frac{\alpha_1+\alpha_3}{\alpha_4}\\\frac{\alpha_1+\alpha_3}{\alpha_4}&\frac{\alpha_2}{\alpha_4})}_T \mqty(v_{2j}\\v_{2j-1}).
\end{equation}

Right away, we observe that $T = i\widetilde{T}$, where $\widetilde{T}$ is a real, symmetric matrix. Further,
\begin{equation}
    \det\qty(T) = \frac{\qty(\alpha_1 + \alpha_3)^2}{\alpha_4^2} + 1 - \frac{\qty(\alpha_1 + \alpha_3)^2}{\alpha_4^2} = 1.
\end{equation}

Thus, the eigenvalues of $T$ are of the form $\qty(i\lambda, \frac{-i}{\lambda}); \lambda\in\mathbb{R}$. Explicitly, the eigenvalues of $T$ are $\lambda_{1,2} = i\frac{\cos(h)\cosh(\beta)}{\sin(h)\sinh(\beta)}, -i\frac{\sin(h)\sinh(\beta)}{\cos(h)\cosh(\beta)}$. Their corresponding eigenvectors are
$\mqty(\alpha_1 \pm 1\\ \alpha_2)$.

We now use the boundary equation for $a_1$ in order to fix the free parameters $\mqty(v_2\\v_1)$. Setting $\mqty(v_2\\v_1) = \mqty(\cos(h)\\ \sin(h)) \propto \mqty(\alpha_1 + 1\\ \alpha_2)$, we find that 
\begin{equation}\label{eq:finres}
\mqty(v_{2j}\\v_{2j-1}) = \lambda_1^{j-1}\mqty(\cos(h)\\ \sin(h)).    
\end{equation}
 Lastly, we require that the edge mode localized around $j=1$ decays exponentially, which imposes that

\begin{equation}
    |\lambda_1|<1 \implies \cosh(2\beta)\cos(2h) < -1, 
\end{equation}

\begin{figure}
    \centering
    \includegraphics[width=0.33\textwidth]{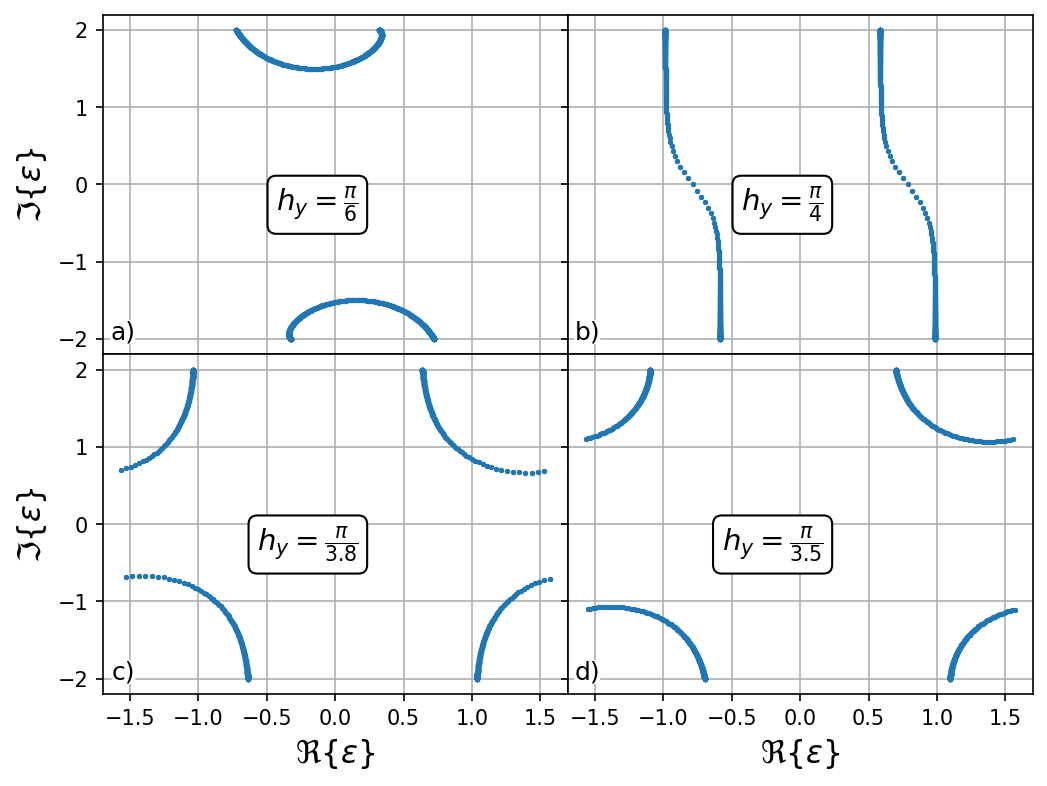}\includegraphics[width=0.33\textwidth]{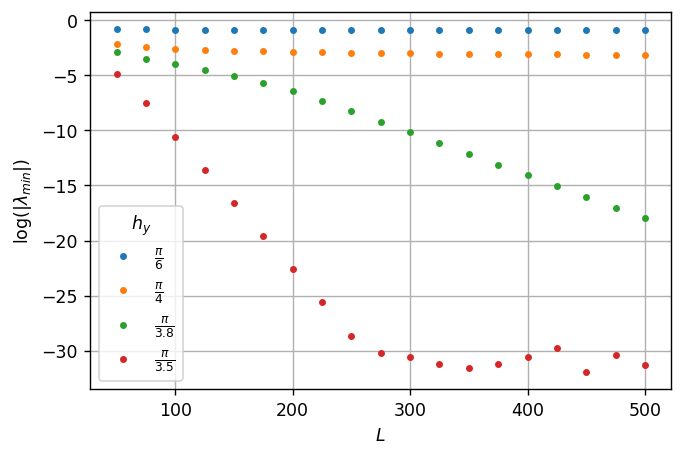}\includegraphics[width=0.33\textwidth]{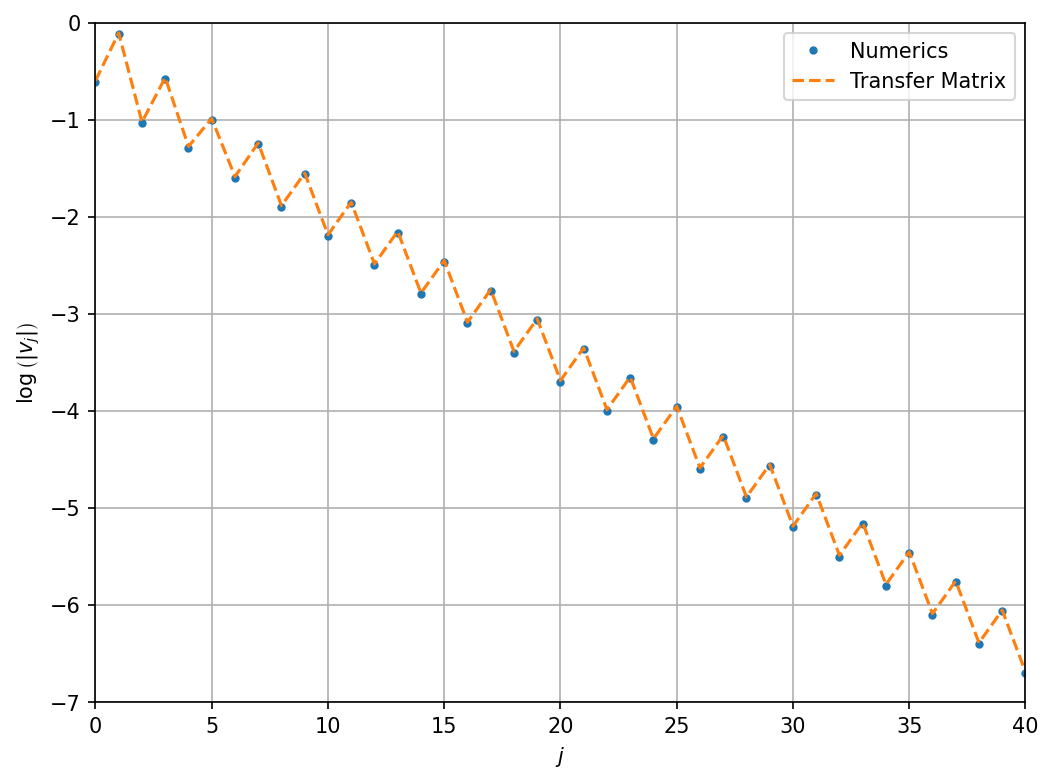}
    \caption{(Left) The spectrum for different values of $h_y$, for $L=1000, J_{xx}=0.2$ and $\beta=2$. (Center) The scaling of the magnitude of the smallest eigenvalue of $T$ against $L$. When the system is in a non-trivial state, the smallest eigenvalue of $M$ decays exponentially with $L$, but not otherwise. (Right) The decay of the $i0$ mode for $L=1000, h_y = \frac{\pi}{3}, J_{xx}=0.2$ and $\beta=2.0$, compared against the eigenvector of $M$ with the smallest magnitude eigenvalue}
    \label{fig:bedge2}
\end{figure}

\begin{figure}
    \hspace{-3em}
        \includegraphics[width=0.62\textwidth]{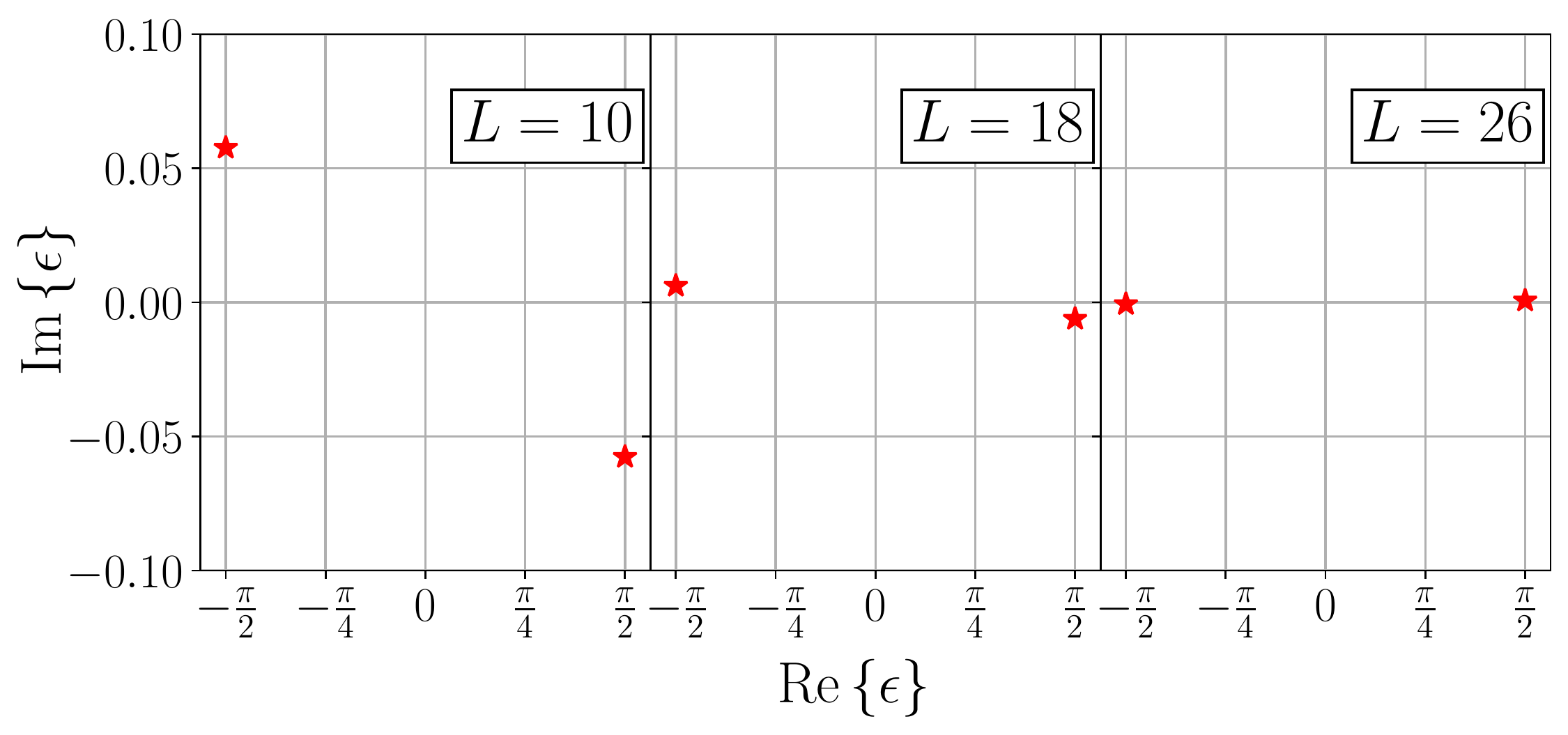}\includegraphics[width=0.38\textwidth]{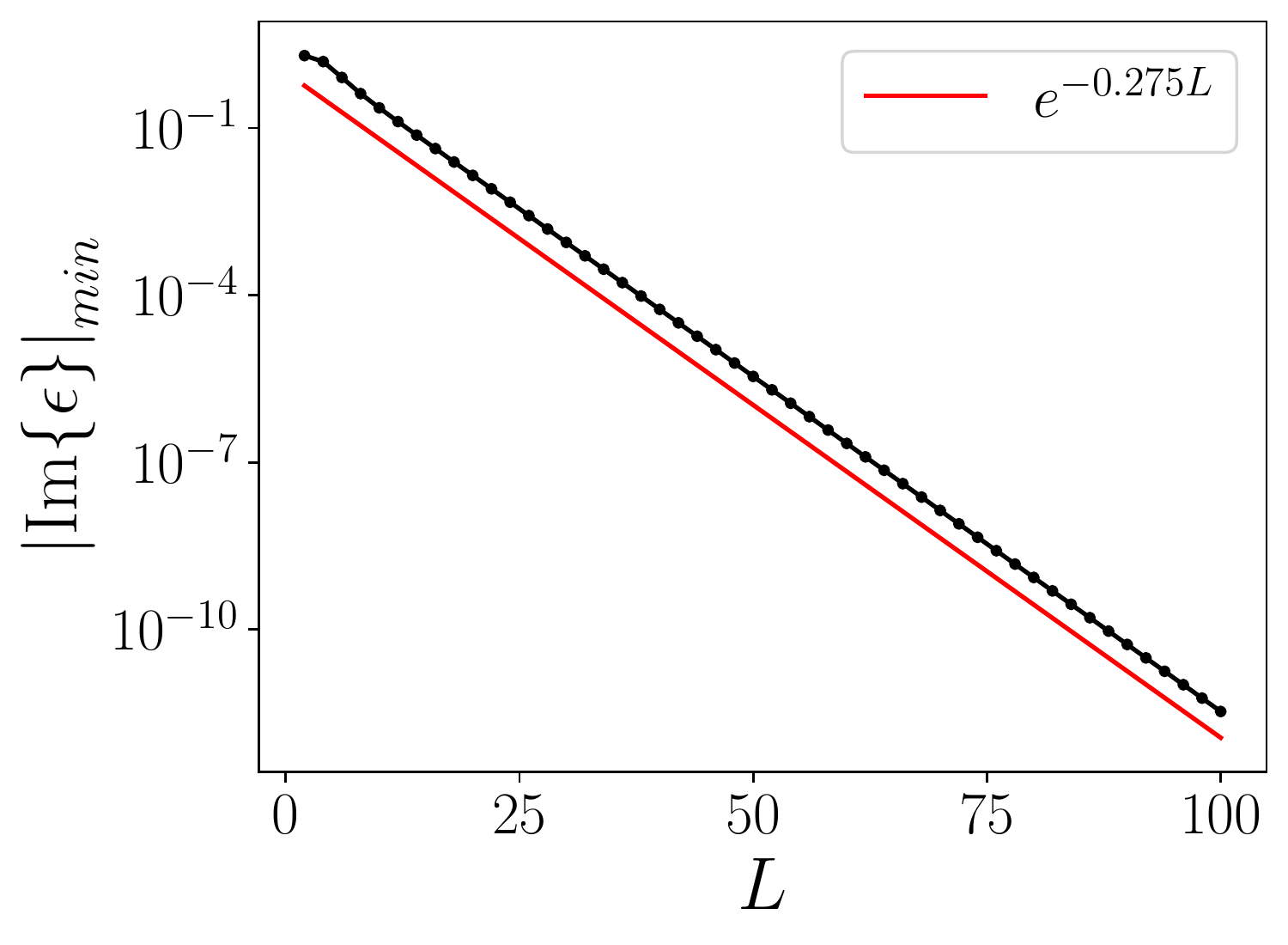}
        \caption{Information about the single-particle spectrum of the model at $\beta=2, h_y=\frac{\pi}{3}$ and $J_{xx} = J_{yy} = J_{zz} = 0$, for different system sizes $L$. (Left) A plot of the portion of the spectrum around $\Im{\epsilon}=0$, showing that there is a splitting in the imaginary direction induced by finite size effects. (Right) The splitting in the imaginary direction, defined as the minimum of the absolute values of imaginary parts of the eigenvalue, decays exponentially with system size $L$. }
        \label{fig:finsize}
    \end{figure}

We can analogously find the expression for the right edge mode by beginning with the boundary equation for $b_L$, and propagating leftwards with $T^{-1}$.

A few other checks were performed to confirm that the operator $F$ obtained through the transfer matrix procedure is, in-fact, the $i0$ modes of interest. We first explicitly construct $\hat{F} = \sum\limits_{j=1}^L v_{2j-1} a_j + v_{2j} b_j$, and represent it as a column vector 

\begin{equation*}
    F_0 = \mqty(v_2\\ v_1\\ v_4\\ v_3 \\ \vdots \\ v_{2L}\\v_{2L-1})
\end{equation*}

so that it corresponds to the representation introduced in \cref{seq:vdefine}. We find that $e^{-i\mathcal{H}_F} F_0 = -F_0$, confirming that $\acomm{\hat{V}}{\hat{F}} = 0$.

While it might not be possible to analytically construct the edge modes for more general $\hat{V}$, it is straightforward to construct the matrix $M$ and thus numerically obtain its smallest magnitude eigenvalue. Again, exponentially small eigenvalues appear exactly as we tune the parameters through a closing of the bulk gap in the imaginary direction. We show this below for $\hat{V} = e^{\beta \sum Z_j Z_{j+1}}e^{-iJ_{xx} \sum X_j X_{j+1}} e^{-i \sum hY_{j}}$.

Finally, we show in \cref{fig:bedge2} that the $i0$ mode obtained by the direct diagonalization of $\hat{V}$ is identical to the mode constructed from the transfer matrix, for nonzero $J_{xx}$.

\section{Time Evolution of Fermionic Gaussian States}
	
	A fermionic many-body state $\ket{\psi}$ is said to be a Gaussian state \cite{Bravyi2005} if  
	\begin{equation}
	\dyad{\psi} = \frac{e^{-\frac{\vec{v}_c^\dagger \mathcal{H}_G^c \vec{v}_c}{2}}}{\tr(e^{-\frac{\vec{v}_c^\dagger \mathcal{H}_G^c \vec{v}_c}{2}})} = \frac{e^{-\frac{\vec{\gamma}^\dagger \mathcal{H}_G \vec{\gamma}}{4}}}{\tr(e^{-\frac{\vec{\gamma}^\dagger \mathcal{H}_G \vec{\gamma}}{4}})}
	\label{seq:Gauss}
	\end{equation} 
	for some $2L\cross 2L$ matrix $\mathcal{H}_G^c = {\mathcal{H}^c_G}^\dagger$ which has the form
	
	\begin{equation}\label{seq:hform}
	\mathcal{H}^c_G = \mqty(A_{L\cross L} & B_{L\cross L} \\B_{L\cross L}^\dagger&-A_{L\cross L}^T)
	\end{equation}
	
	with $A^\dagger = A$ and $B^T = -B$. For use in this section, we have defined $\vec{v}_c$ analogously to $\vec{\gamma}$ as
	
	\begin{equation}
	    \label{eq:vc_def}
	    \vec{v}_c = \mqty(c_1\\ c_2 \\ \vdots\\ c_L\\c_1^\dagger \\ \vdots \\c_L^\dagger),
	\end{equation}
	
	where $\qty{c_j,c_j^\dagger}_{j=1}^L$ denote complex fermionic annihilation and creation operators at site $j$, obeying 
	$\qty{c_j,c^\dagger_k}=\delta_{j,k}$. These are related to $\qty{\gamma_j}$ as
	
	\begin{equation}
	    \begin{aligned}
	    \gamma_{2j-1} &= i(c_j-c_j^\dagger)\\
	    \gamma_{2j} &= (c_j + c_j^\dagger)
	    \end{aligned}
	\end{equation}
	
	The relationship between $\vec{\gamma}$ and $\vec{v}_c$, and $\mathcal{H}_G$ and $\mathcal{H}^c_G$ can be succinctly expressed through the $2L\cross2L$ matrix $W$
	
	\begin{equation}
	    W = \mqty(i&0& \cdots& -i&0&\cdots\\
	    1&0&\cdots&1&0&\cdots\\
	    0&i&\cdots&0&-i&\cdots\\
	    0&1&\cdots&0&1&\cdots\\
	    &&&\vdots&\\
	    0&\cdots&1&0&\cdots&1\\)
	\end{equation}
	
	as 
	
    \begin{equation}
        \vec{\gamma} = W\vec{v}_c
    \end{equation}

	and
	
	\begin{equation}
        \mathcal{H}^c_G = \frac{1}{2} W^\dagger\mathcal{H}_GW.
    \end{equation}
	
	$\frac{W}{\sqrt{2}}$ is a unitary matrix obeying $WW^\dagger=W^\dagger W=2\mathbb{1}$, so $\mathcal{H}_G^c$ and $\mathcal{H}_G$ are unitarily similar and thus, share an eigenspectrum, as expected. For a quadratic fermionic Hamiltonian $\hat{H}$, this relation generally holds between the  $\mathcal{H}$ and $\mathcal{H}^c$, the matrices that represent $\hat{H}$ in terms of majorana and complex fermions, respectively.
	
	A Gaussian state obeys Wick's theorem, and every state that obeys Wick's Theorem can be expressed as in \cref{seq:Gauss}. The expectation value in this state of any $N$-body operator can be written in terms of contractions involving the 2-point correlator $\expval{\gamma_i \gamma_j}$. Such states can be completely characterized in terms of their correlation matrices (or $C-$ matrices)
	\begin{equation}
	\begin{aligned}
	C^m_{i,j} = \expval{\gamma_i \gamma_j}\\
	C^c_{i,j} = \expval{\vec{v}^{}_{ci} \vec{v}^\dagger_{cj}}
	\end{aligned}
	\end{equation}
	
	where $c(m)$ denote the expectation values of complex (majorana) fermions. The two matrices are related as
	\begin{equation}\label{crel}
	C^m = W C^c W^\dagger
	\end{equation}.

	and 

	\begin{equation}\label{seq:chrel}
	\begin{aligned}
	C^c &= (1 + e^{-\mathcal{H}_G^c})^{-1},\\
	C^m &= 2(1 + e^{-\mathcal{H}_G})^{-1}.
	\end{aligned}
	\end{equation}

	The description of states that are defined on a $2^L$ dimensional Hilbert space has now been reduced to a description only involving $2L$ dimensions, avoiding the exponential growth of computational resources. Thus, instead of tracking the evolution of the state $\ket{\psi}$, it suffices to study its corresponding $C-$ matrix, provided that the state under consideration always remains Gaussian.
	
	\subsection{Unitary Time Evolution of the C-Matrix}
	
	Under a time evolution governed by any fermionic Hamiltonian that is quadratic in the creation/annihilation operators, an initial Gaussian state always remains a Gaussian state. So, for some Hamiltonian $\hat{H} =\frac{\vec{v}_c^\dagger \mathcal{H}^c \vec{v}_c}{2}$, the time evolution of the $C-$ Matrix is given as 
	
	\begin{equation}\label{seq:timeevol1}
	\begin{aligned}
	C^c (t) = \mathcal{U} C^c \mathcal{U}^\dagger\\
	\mathcal{U}(t) \equiv e^{-i\mathcal{H}t}
	\end{aligned}
	\end{equation}

	Owing to the product rule \cref{seq:prodrule}, the extension to Floquet systems proceeds using the Floquet Hamiltonian $\mathcal{U} = e^{- i \mathcal{H}^c_F}$ in \cref{seq:timeevol1}. 
	
	\subsection{Non-Unitary Evolution - Establishing Gaussianity}
	
	It is not immediately obvious that $\hat{V}$ maps one Gaussian state to another. In order to show this, we must make use of the product rule in \cref{seq:prodrule}. Generalizing the evolution rule to density matrices, we have
	\begin{equation}\label{seq:erule2}
	\rho \to \frac{V\rho V^\dagger}{\tr(V\rho V^\dagger)},
	\end{equation}
	
	where $\rho$ is the density matrix $\dyad{\psi}$ corresponding to the pure state $\ket{\psi}$.
	
	As with the time evolution of pure states, it is this explicit normalization that makes this time evolution non-linear. In the majorana representation, $\hat{V} = e^{-i\frac{\vec{\gamma}^\dagger \mathcal{H}_F \vec{\gamma}}{4}}$. Defining $\mathcal{V} = e^{-i\mathcal{H}_F}$, we have, by the product rule, 
	
	\begin{equation}
	\begin{aligned}
	e^{-\frac{\vec{\gamma}^\dagger \mathcal{H}_G \vec{\gamma}}{4}} \to e^{-\frac{\vec{\gamma}^\dagger \mathcal{H}'_G \vec{\gamma}}{4}}\\
	e^{-\mathcal{H}'_G} = \mathcal{V}e^{-\mathcal{H}_G} \mathcal{V}^\dagger
	\end{aligned}
	\end{equation}
	
	Since $\mathcal{H}_F$ and $\mathcal{H}_G$ are antisymmetric, so is $\mathcal{H}'_G$. Further, the Hermiticity of $\mathcal{H}_G$ ensures the Hermiticity of $\mathcal{H}'_G$. The resulting state $\rho' = 	\frac{e^{-\frac{\vec{\gamma}^\dagger \mathcal{H}'_G \vec{\gamma}}{4}}}{\tr(e^{-\frac{\vec{\gamma}^\dagger \mathcal{H}'_G \vec{\gamma}}{4}})} $ is also a Gaussian state. This proof relied only on the quadratic nature of $\mathcal{H}_F$ and its antisymmetry, making it equally valid for unitary evolution. With the existence of a corresponding $C-$ matrix $C'$ ensured, we now focus on relating $C'$ to $C$.
	
	\subsection{Equations of Motion for $C$}
	
The first method provides an equation of motion for the $C-$ matrix, under the assumption that $\hat{V}$ can be decomposed into a real time 
and imaginary time 
evolution. Unitary evolution is implemented directly by \cref{seq:timeevol1}. Then, using the fact that $C = (1 + e^{-\mathcal{H}^c_G})^{-1}$, defining an intermediate $\widetilde{C}'(x) = (1 + e^{-x \mathcal{H}^c_I}e^{-\mathcal{H}^c_G}e^{-x \mathcal{H}^c_I})^{-1}$ and differentiating w.r.t $x$, we have
	
	\begin{equation}
	\label{seq:timeevol2}
	\dv{C}{x} = \acomm{H_I}{C} + 2 C H_I C
	\end{equation}
	
	which can be numerically integrated to give the $C-$ matrix.
	
	\subsection{Mapping Annihilation Operators}
	
	A more efficient approach is to directly construct the $C-$ matrix at every time step. This process begins by noting that a pure state in this enlarged $2L$ dimensional space is \textit{always} at \lq\lq{}half-filling\rq\rq{}, i.e. $\tr(C) = L$. This means that for any state, there exist exactly $L$ operators that annihilate it. For example, consider a 1-D chain which has the first $N$ sites occupied 
	
	\begin{equation}
	\begin{aligned}
	\ket{\psi} &= \prod\limits_{j=1}^N c^\dagger_j\ket{0},\\
	c_j\ket{\psi} &= 0, j=N+1 \cdots L,\\
	c^\dagger_j\ket{\psi} &= 0, j=1 \cdots N.
	\end{aligned}
	\end{equation}
	
	Thus, $\ket{\psi}$ can be thought of as either $N$-filled sites, starting from the vacuum, or $L-N$ particles removed from the fully occupied chain. One finds that by simply keeping track of the operators that annihilate the state at a given time, the entire $C-$ matrix can be recreated. In the basis of the operators that annihilate a state (represented as $\vec{d}$), with $d_j \ket{\psi} = 0; j=1\cdots L$, the only non-zero elements of the $C-$ matrix are $\expval{d_j d^\dagger_j} = 1$. That is, in this basis,
	\begin{equation}\label{C0}
	C = \mqty(\mathbb{1}_L&0_L \\ 0_L& 0_L) \equiv C_0.
	\end{equation}
	
	Secondly, for two bases $\lbrace c_j, c^\dagger_j\rbrace$ and $\lbrace d_j, d^\dagger_j\rbrace$ related by a unitary transformation $\vec{v}_c = \mathcal{U} \vec{v}_d$, the two $C-$ matrices are related by 
	\begin{equation}\label{Crel}
	\begin{aligned}
	C = \mathcal{U} C_0\text{ }\mathcal{U}^\dagger;\\
	C_{ij} = \sum\limits_{l=1}^L \mathcal{U}_{i,l}. \mathcal{U}^\dagger_{l,j}
	\end{aligned}
	\end{equation}
	
	The transformation $\mathcal{U}$ is guaranteed to be unitary since it is both linear and canonical (preserves commutation relations).
	
	\subsection{Finding $\mathcal{U}$}
	
	Consider an initial state $\ket{\psi_0}$ and a set of creation and annihilation operators $\lbrace c_j, c^\dagger_j\rbrace$ such that $c_j\ket{\psi_0} = 0, j = 1\cdots L$. The evolution of this state is given by
	\begin{equation}
	\begin{aligned}
		\ket{\psi_1} = \frac{\hat{V}\ket{\psi_0}}{\norm{\hat{V}\ket{\psi_0}}};\\
		d_j \ket{\psi_1} = 0, j=1\cdots L.
	\end{aligned}
	\end{equation}
	We now wish to find $\mathcal{U}$ that relates $\lbrace c_j\rbrace$ and $\lbrace d_j\rbrace$. Consider
	\begin{equation}
	\begin{aligned}
		\tilde{d}_j \equiv \hat{V}c_j\hat{V}^{-1};\\
		\tilde{d}_j \ket{\psi_1} \propto \hat{V}c_j\hat{V}^{-1}\hat{V}\ket{\psi_0} = 0.
	\end{aligned}
	\end{equation}
	
	Right away,
	\begin{equation}\label{acommd}
	\acomm{\tilde{d}_j}{\tilde{d}_l} = \acomm{\hat{V}c_j\hat{V}^{-1}}{\hat{V}c_l\hat{V}^{-1}} = \hat{V}\acomm{c_j}{c_l}\hat{V}^{-1} = 0.
	\end{equation}
	
	Since $\hat{V}$ is Gaussian, 
	\[\tilde{d}_j = \sum\limits_{l=1}^L \left(\mathcal{V}_{j,l} c_l + \mathcal{V}_{j,l+L} c^\dagger_l\right).\]
	
	However, in general, $\acomm{\tilde{d}^\dagger_j}{\tilde{d}_l} \neq \delta_{jl}$. This is resolved as follows.
	
	We begin by writing each of the $L$ annihilation operators as linear combinations of the canonical $\lbrace c_j, c^\dagger_j\rbrace$ operators. An arbitrary operator $d_j$ can be written as 
	
	\begin{equation}
	d_i = \vec{\alpha}_i^\dagger \vec{v}_c.
	\end{equation}
	
	Instead of attempting to study operator evolution directly, we can instead focus on the $2L$ dimensional complex vector $\vec{\alpha}_i$. Under the evolution rule $c_j \to \hat{V} c_j \hat{V}^{-1} = \vec{\alpha}_i^\dagger\mathcal{V}^{-1}\vec{v}_c$, we can instead define the evolution as being generated by
	
	\begin{equation}\label{seq:evolRuleAB}
	\vec{\alpha}_i \to \vec{\tilde{\beta}}_i = \left(\mathcal{V}^{-1}\right)^\dagger\vec{\alpha}_i.
	\end{equation}
	
	Owing to the Gaussian nature of the time evolution, we can collect the $L$ vectors corresponding to the $L$ annihilation operators $\left\lbrace c_j \right\rbrace_{j=1}^L$ in a $2L\cross L$ matrix, which we shall call $\mathcal{U}_0$.
	
	\begin{equation}
	\mathcal{U}_0 \equiv \begin{pmatrix}
	\vert & \vert & &\vert\\
	 \vec{\alpha}_1  & \vec{\alpha}_2 &\cdots& \vec{\alpha}_L\\
	\vert & \vert& &\vert
	\end{pmatrix}_{2L\cross L}.
	\end{equation}
	
	By \cref{seq:evolRuleAB}, we have
	\begin{equation}
	\mathcal{U}_0 \to \widetilde{\mathcal{U}}_1 = \begin{pmatrix}
	\vert & \vert & &\vert\\
	\vec{\tilde{\beta}}_1  & \vec{\tilde{\beta}}_2 &\cdots& \vec{\tilde{\beta}}_L\\
	\vert & \vert& &\vert
	\end{pmatrix}.
	\end{equation}
	
	The operators given by $\tilde{d}_i \equiv \vec{\tilde{\beta}}_i^\dagger \vec{v}_c$ do not obey canonical commutation relations with their Hermitian conjugates. However, since $d_j \ket{\psi_1} = 0$, this is also true for any linear combination of the $\left\lbrace d_j \right\rbrace$. This implies that any vector constructed as a linear combination of $\left\lbrace \vec{\tilde{\beta}}_i \right\rbrace$ also describes an operator that annihilates $\ket{\psi_0}$. Additionally, since $\left\lbrace \tilde{d}_j \right\rbrace$ anticommute with each other, so too do their linear combinations. Therefore, we consider the vectors $\left\lbrace \vec{\beta}_j \right\rbrace$ obtained by orthonormalizing $\left\lbrace \vec{\tilde{\beta}}_j \right\rbrace$, collected in the matrix
	
	\begin{equation}
	\mathcal{U}_1 = \begin{pmatrix}
	\vert & \vert & &\vert\\
	\vec{\beta}_1  & \vec{\beta}_2 &\cdots& \vec{\beta}_L\\
	\vert & \vert& &\vert
	\end{pmatrix}.
	\end{equation}
	
	The operators $\left\lbrace d_j = \vec{\beta_j}^\dagger\vec{v}_c \right\rbrace$ are canonical. We already showed in \cref{acommd} that they anticommute amongst themselves. Further,
	\begin{equation}
	\begin{aligned}
	\acomm{d_i}{d^\dagger_j}& = \sum\limits_{k,l=1}^L\acomm{(\beta_i)^*_k c_k + (\beta_i)^*_{k+L} c^\dagger_k}{(\beta_j)_l c^\dagger_l + (\beta_j)_{l+L} c_l} \\
	& = \sum\limits_{k,l=1}^L\acomm{(\beta_i)^*_k c_k }{(\beta_j)_l c^\dagger_l} + \acomm{(\beta_i)^*_{k+L} c^\dagger_k}{(\beta_j)_{l+L} c_l} \\
	& = \sum\limits_{k=1}^{2L} (\beta_i)^*_k (\beta_j)_k = \delta_{ij}. \hspace{2em}\left[\text{By orthnormalization}\right]
	\end{aligned}
	\end{equation}

	The resulting $L\cross 2L$ matrix $\mathcal{U}^\dagger_1$ which relates $\lbrace d_j\rbrace_{j=1}^L$ to 
	$\lbrace c_j, c^\dagger_j\rbrace$ is exactly the part of $\mathcal{U}^\dagger$ that we require from \cref{Crel}. Finally, we have
	
	\begin{equation}\label{fineq}
	C = \mathcal{U}_1\mathcal{U}^\dagger_1.
	\end{equation}

\section{Gapless Phase and Transitions by Tuning $\beta$}

\begin{figure}
    \centering
    \includegraphics[width=0.5\textwidth]{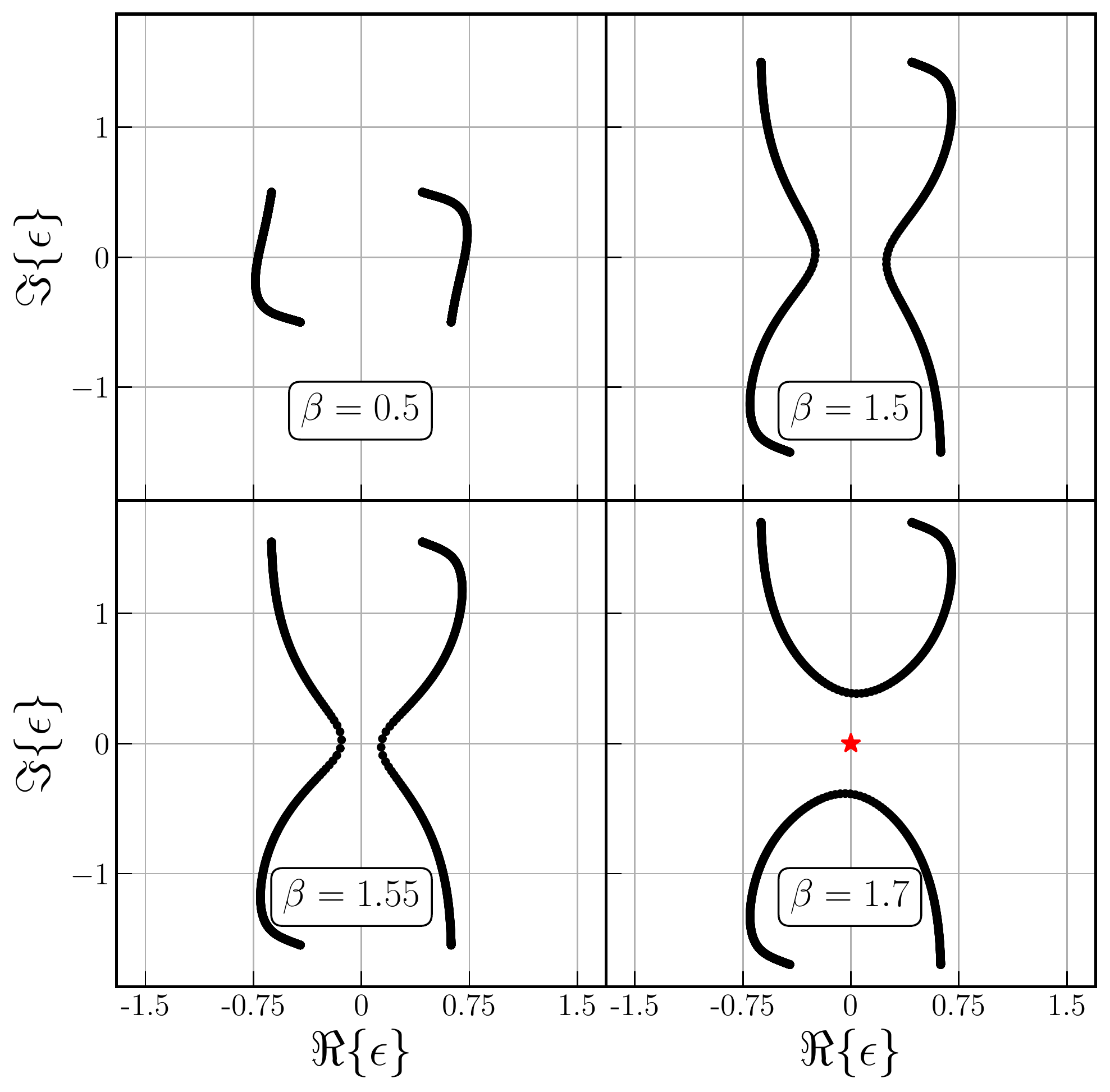}%
    \includegraphics[width=0.5\textwidth]{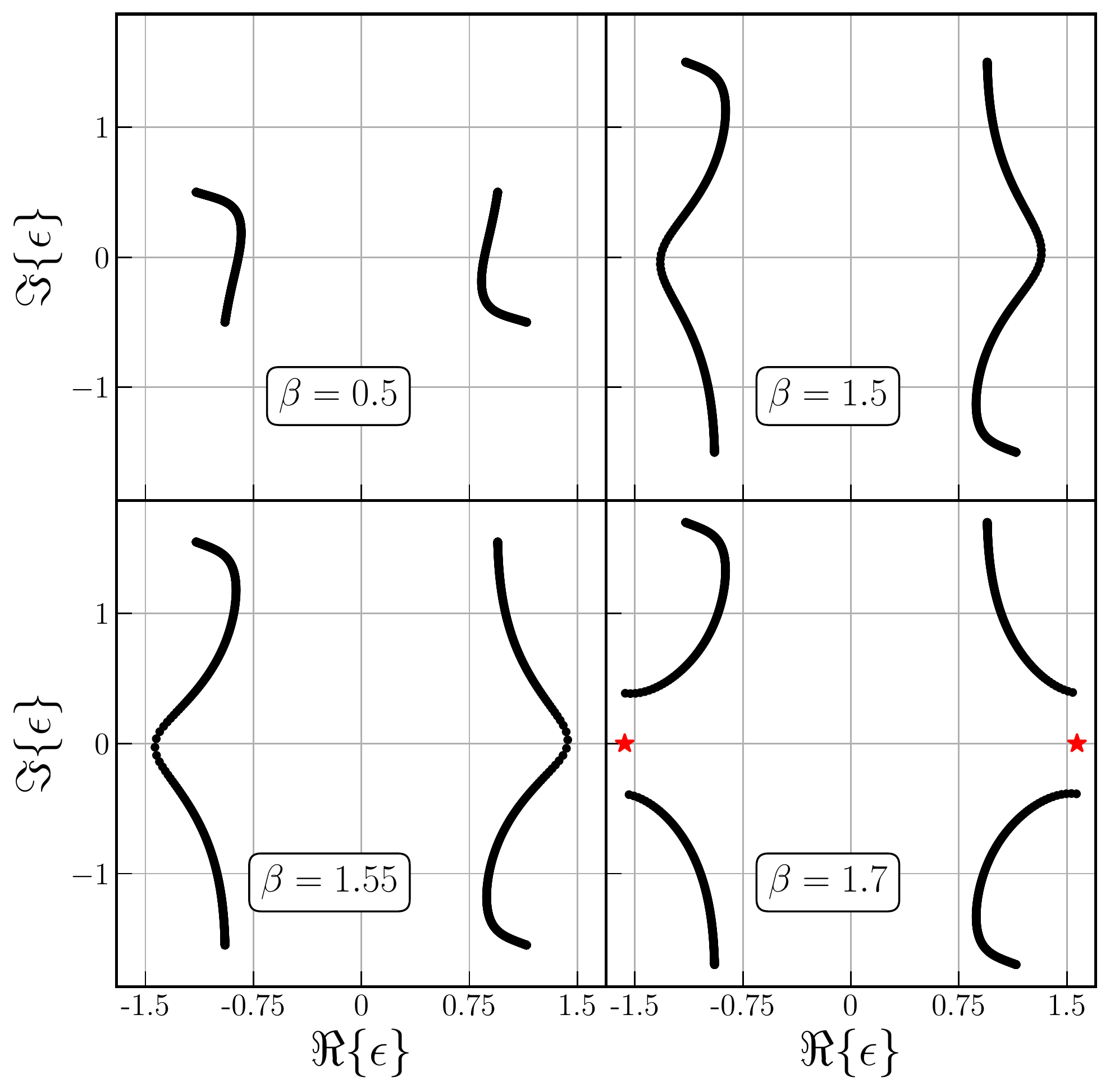}
    \caption{Plots of the complex spectrum showing an imaginary gap opening as $\beta$ is increased, revealing $i0$ modes (Left) with 0 splitting, $h_y = \frac{\pi}{6}$ (Right) with a $\pi$ splitting, $h_y=\frac{\pi}{3}$ in the real direction.}
    \label{fig:spect-gapless}
\end{figure}

\begin{figure}
    \centering
    \includegraphics[width=0.5\textwidth]{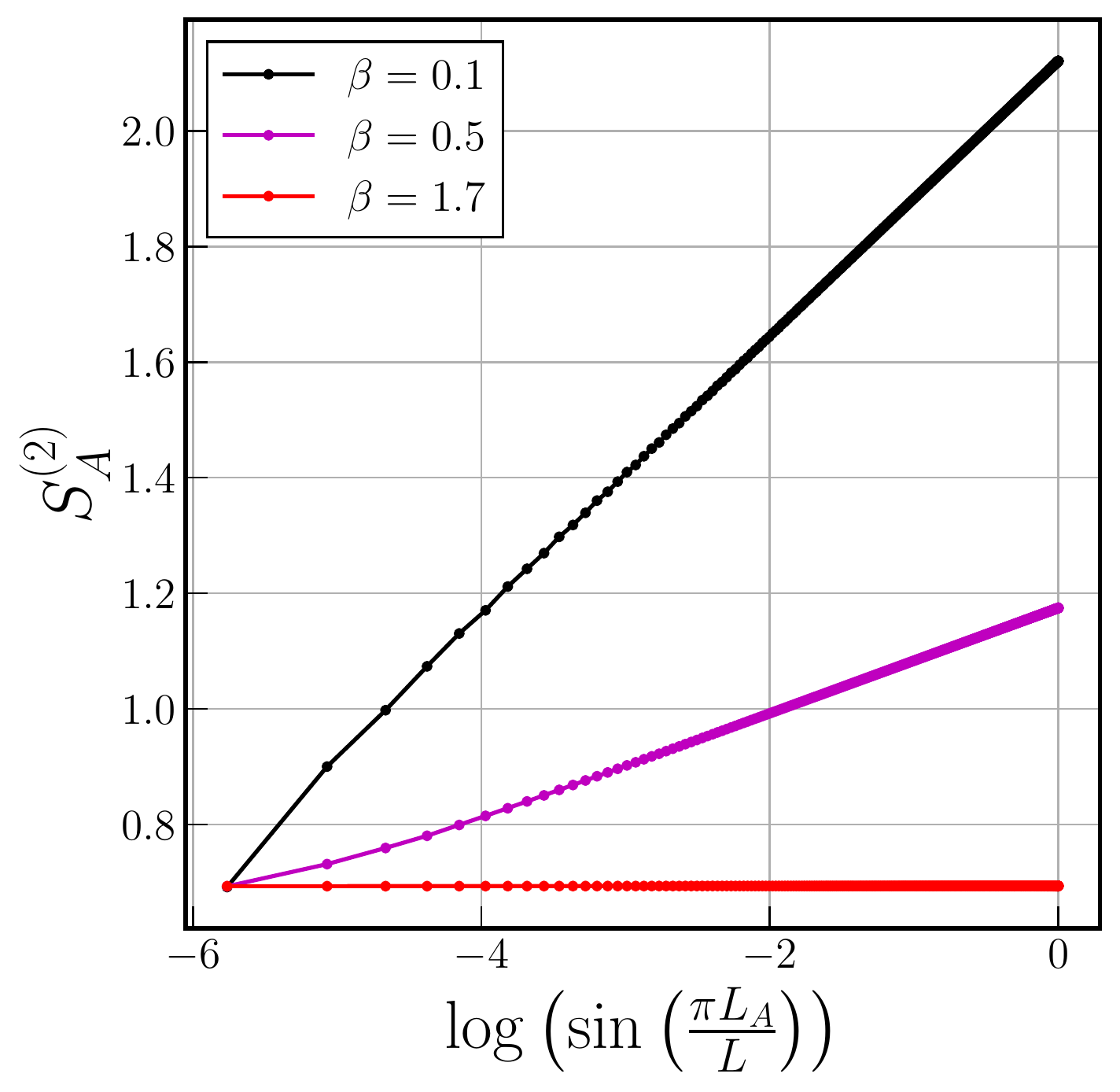}%
    \includegraphics[width=0.5\textwidth]{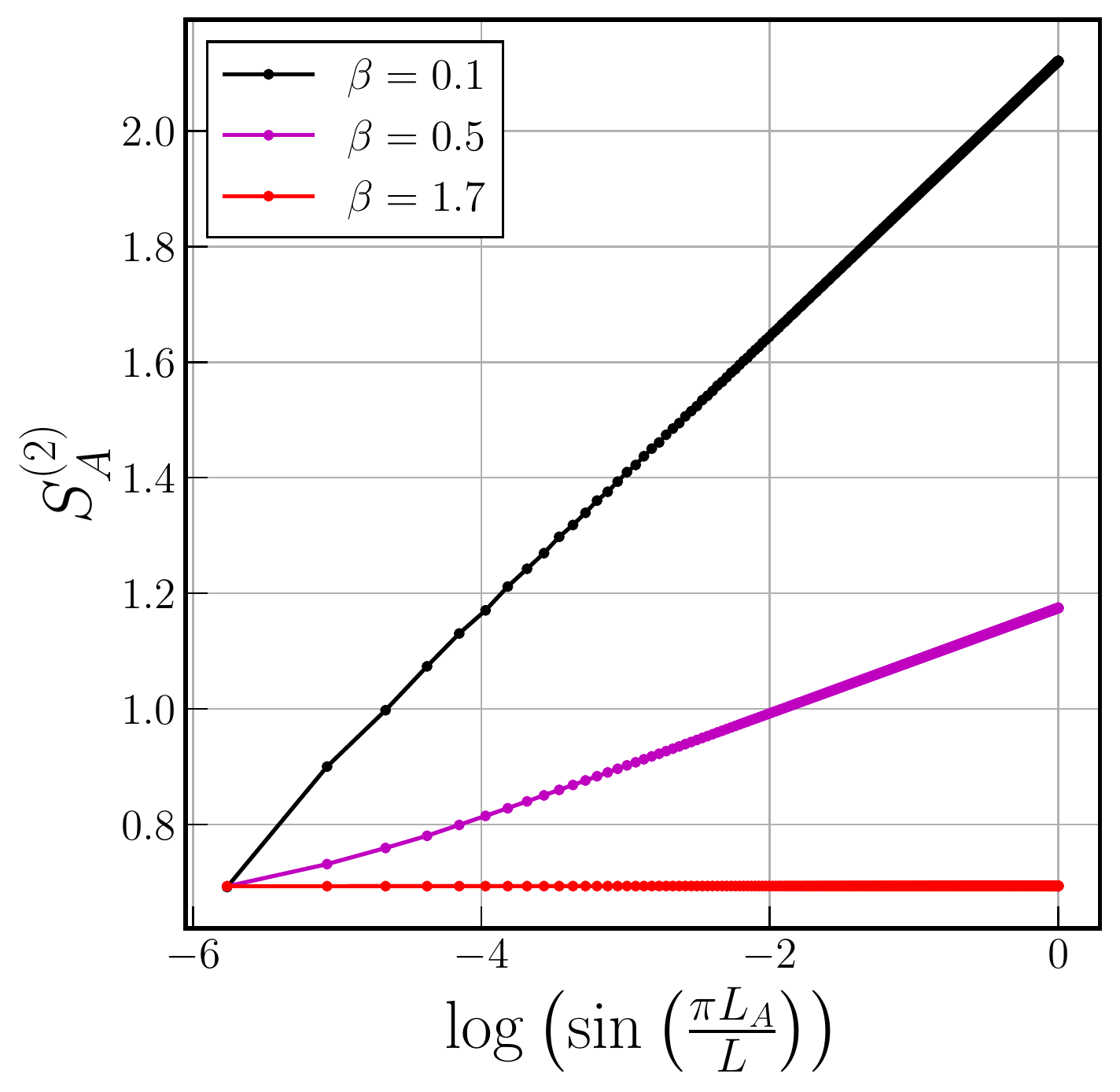} 
    \caption{Scaling of the entanglement entropy with subsystem size as $\beta$ is increased. (Left) $h_y = \frac{\pi}{6}$ (Right) $h_y=\frac{\pi}{3}$. When the $i0$ modes are present, the entanglement entropy becomes independent of system size.}
    \label{fig:ent-gapless}
\end{figure}

Lastly, we turn to the phase diagram as $\beta$ is varied. When $h_y$ is set to be appreciably close to $\frac{\pi}{2}$, we observe an imaginary gap open as $\beta$ increases, leaving 2 $i0$ modes with a $\pi$ splitting between their real parts. For $h_y$ closer to zero, the $i0$ modes are degenerate instead.

As might be expected for these gapless modes, when the entanglement entropy of the steady state is considered, there is a transition from a critical, logarithmic to an area law scaling with the subsystem size. The coefficient of $\log(\sin(\frac{\pi L_A}{L}))$ in the critical phase is parameter dependent, in agreement with previous results on emergent conformal symmetry in non-unitary random free fermion models \cite{Chen_2020,Alberton_2021,jian2020criticality}.

\bibliography{suppbiblio}